\documentclass[twocolumn,showpacsfAppendix,prd,nofootinbib,superscriptaddress,amsmath,amssymb,floatfix]{revtex4-1}
\usepackage{graphicx,color}  
\usepackage{dcolumn}   
\usepackage{bm}        
\usepackage{amssymb}   
\usepackage{lineno, soul, ulem}

\newcommand{\BibitemShut}[1]{}

\def \aap  {A\&A}

\def \apj  {ApJ}
\def \apjs  {ApJS}

\def \jcap  {JCAP}

\def \prl {PRL}

\begin{document}

\preprint{TTK-XX-XX}


\title{Novel interpretation of the latest AMS-02 cosmic-ray electron spectrum}


\author{Mattia Di Mauro}
\affiliation{Istituto Nazionale di Fisica Nucleare, via P. Giuria, 1, 10125 Torino, Italy}
\author{Fiorenza Donato}
\affiliation{Department of Physics, University of Torino, via P. Giuria, 1, 10125 Torino, Italy}
\affiliation{Istituto Nazionale di Fisica Nucleare, via P. Giuria, 1, 10125 Torino, Italy}
\author{Silvia Manconi}
\affiliation{Institute for Theoretical Particle Physics and Cosmology, RWTH Aachen University, Sommerfeldstr.\ 16, 52056 Aachen, Germany}

\date{\today}

\begin{abstract}
The latest AMS-02 data on cosmic ray electrons show a break in the energy spectrum around 40~GeV, with a change in the slope of about 0.1. 
We perform a combined fit to the newest AMS-02 positron and electron flux data above 10 GeV using a semi-analytical diffusion model where sources includes production of pairs from pulsar wind nebulae (PWNe), electrons from supernova remnants (SNRs) and both species from spallation of hadronic cosmic rays with interstellar medium atoms.
We demonstrate that within our setup the change of slope in the AMS-02 electron data is well explained by the interplay between the flux contributions from SNRs and from PWNe. In fact, the relative contribution to the data of these two populations changes by a factor of about 13 from 10 to 1000~GeV.
The PWN contribution has a significance of at least $4\sigma$, depending on the model used for the propagation, interstellar radiation field and energy losses. We checked the stability of this result against low-energy effects by solving numerically the transport equation. as well as adding possible breaks in the injection spectrum of SNRs. 
The effect of the energy losses alone, when the inverse Compton scattering is properly computed within a fully numerical treatment of the Klein-Nishina cross section, cannot explain the break in the $e^-$ flux data, as recently proposed in the literature. 
\end{abstract}

\maketitle


\section{Introduction} 

The Alpha Magnetic Spectrometer (AMS-02) is a state-of-the-art particle physics detector operating on the International Space Station.
By taking data since 2011, it is providing very precise measurements of cosmic-ray (CR) fluxes for leptons and nuclei, from hydrogen up to silicon, as well as for rare antiparticles such as positrons and antiprotons.
Of all the CR species, electrons ($e^-$) and positrons ($e^+$) are among the most intriguing ones to study.
In fact, they are probably produced by the superposition of different Galactic sources and physical production mechanisms which, due to the intense radiative losses suffered, test the properties of our Galactic environment within a few kpc. 

Very recently, AMS-02 published new separate spectra for the $e^{+}$ and $e^{-}$ fluxes, reaching the unprecedented energy of 830 GeV for the former and 1.2 TeV for the latter \cite{PhysRevLett.122.041102,PhysRevLett.122.101101}.
The $e^-$ data show a significant excess above about 42 GeV when compared to the trend of the spectrum at lower energies. Interestingly, the nature of this excess is different from the excess in the $e^+$ flux detected above 25.2 GeV which has an exponential energy cutoff at about 800 GeV.
The $e^-$ data can be well fitted by a smooth broken power law with a break at about 42 GeV and a difference of slope between the low and high-energy power laws $\Delta \gamma \approx 0.1$. By fitting the AMS-02 $e^-$ data above 10 GeV with a single power-law and with a broken power-law function we find that the model with the break is preferred at about $5\sigma$ significance. We perform a fit to the data above 10 GeV using a power-law and a broken power-law functions finding a $\chi^2$ of 39.0 and 8.7, respectively. Therefore, the $\Delta \chi^2$ between the two cases is 30.3 that, considering the two additional parameters of the broken power-law (the break and spectral index above the break), gives a significance of $5.0$ for the presence of a break in the AMS-02 $e^-$ spectrum. We do not include in the calculation any systematic uncertainty due to energy measurement.
Previously, the Fermi Large Area Telescope ({\it Fermi}-LAT) Collaboration published the $e^+ + e^-$ inclusive spectrum from 7 GeV to 2 TeV, and reported the detection of a break at similar energies $\sim 50$ GeV  with $\Delta \gamma $ compatible with the one measured by AMS-02 \cite{Abdollahi:2017nat}. The significance of the spectral break is at the $4\sigma$ level if the uncertainties on the energy measurements are taken into account \cite{Abdollahi:2017nat}.

In Ref.~\cite{DiMauro:2017jpu} the authors explained the {\it Fermi}-LAT data on the inclusive spectrum in terms of $e^-$ produced by supernova remnants (SNRs) through diffusive shock acceleration, $e^{\pm}$ injected in the Galaxy by pulsar wind nebulae (PWNe), and by secondary $e^{\pm}$ produced in inelastic scatterings of hadronic CRs off atoms of the interstellar medium (ISM).
Specifically, the hinted break in the {\it Fermi}-LAT $e^\pm$ flux was well reproduced by the interplay between the SNR and the PWN components, with the latter source population emerging significantly above a few hundreds of GeV.
Previous AMS-02 measurements of separated $e^+$ and $e^-$  \cite{2014PhRvL.113l1102A}, positron fraction \cite{2014PhRvL.113l1101A} and total $e^++e^-$ \cite{Aguilar:2014fea} were studied \cite{DiMauro:2014iia,Lin:2014vja} on their whole energy spectrum. While Ref.~\cite{DiMauro:2014iia} explained the data with a model very similar to the one in \cite{DiMauro:2017jpu}, Ref.~\cite{Lin:2014vja} found that two breaks in the primary $e^-$ injection spectrum - one at about 2.5 GeV and another around 60 GeV -  are needed to explain the data set.

A recent work~\cite{Evoli:2020ash} used a similar model with PWNe, SNRs and secondary production to interpret the latest AMS-02 leptonic data. 
They find that the observed break in the $e^-$ spectrum is due to energy losses for inverse Compton scattering (ICS) suffered by $e^-$ CRs interacting with the photons of the interstellar radiation field (ISRF).
In particular, the change of slope in the data is interpreted as due to the difference between the Thomson regime and the Klein–Nishina formalism for the ICS energy losses, the former valid if $E \epsilon \ll m_e^2 c^4$, where $m_e$ is the electron mass, $E$  the electron energy and $\epsilon$  the photon energy in the lab frame. 
The ISRF is composed by the Cosmic Microwave Background (CMB), and by Galactic dust emission and starlight photons which have a peak in their spectrum at about $\epsilon \approx 2.3\times 10^{-4}/2.4\times 10^{-3}/1.0$ eV, respectively \cite{Vernetto:2016alq}.
The observed energy break in the $e^-$ data is ascribed to the transition of the ICS losses from the Thomson regime to the Klein-Nishina formalism occurring for the starlight, at an interstellar electron energy $m_e^2 c^4/(2 \epsilon) \approx 60$ GeV \footnote{Typically, an $e^-$ produced in the Galaxy at $E= 60$ GeV loses some 10-20\% of its energy during the propagation. Therefore, an interstellar energy of 60 GeV is roughly compatible with the observed energy of the break in the AMS-02 $e^-$ data.}.
Ref.~\cite{Fang:2020dmi} showed that the results of Ref.~\cite{Evoli:2020ash} can be driven by the approximation they used for the ICS loss rate taken from Ref.~\cite{2010NJPh...12c3044S}.
The same authors of Ref.~\cite{Evoli:2020ash} used the more accurate approximation suggested in Ref.~\cite{Fang:2020dmi} to calculate the Klein-Nishina energy losses \cite{Evoli:2020szd}.
Nevertheless, they still interpret the break in the $e^-$ flux with the transition of the ICS energy losses from the Thomson to the Klein-Nishina formalism, though this requires a substantial modification of the starlight ISRF that is increased by a factor of two with respect to what done in their previous paper \cite{Evoli:2020szd}.

The goal of this paper is to find an interpretation of the latest $e^{\pm}$ data, and in particular to assess if the break in the $e^-$ AMS-02 data at about 40 GeV is explained by an energy loss effect as found in Refs.~\cite{Evoli:2020ash,Evoli:2020szd}, or by the interplay between the emission of different source populations, similarly to what found in \cite{DiMauro:2017jpu} by fitting the $e^++e^-${\it Fermi}-LAT data.
We provide the results by testing the ISRF models of Refs.~\cite{Porter_2006,2010A&A...524A..51D,Evoli:2020szd}, by using whether a smooth spatial Galactic distribution of PWNe and SNRs or by including Galactic spiral arms as in \cite{1992ApJS...83..111W}, and by using the propagation parameters in Refs.~\cite{Genolini:2015cta,Evoli:2020ash}.
We calculate, for the first time, the significance for the contribution of PWNe to the measured AMS-02 $e^-$ flux and we
 statistically assess the nature of the break in the data.
 
The paper is organized as follows.
In Sec.~\ref{sec:model} we report the model we use for the acceleration and propagation of CR $e^{\pm}$ from PWNe, SNRs and secondary production. In Sec.~\ref{sec:modellosses} we study the energy losses for ICS on the ISRF. In Sec.~\ref{sec:results} we report our results for the interpretation of the $e^{\pm}$ AMS-02 data. Finally, in Sec.~\ref{sec:conclusions} we write our conclusions.

\section{$e^\pm$ production and propagation in the Galaxy} 
\label{sec:model}
We employ the model already described in Refs.~\cite{DiMauro:2014iia,DiMauro:2017jpu,Manconi:2020ipm} (to which we refer for further details) which assumes $e^{\pm}$ produced by acceleration of $e^-$ from SNRs, pair emission from PWNe and by secondary production in the ISM. Specifically, we use the secondary production calculated in Ref.~\cite{DiMauro:2017jpu} by implementing the primary proton and helium spectra fitted on AMS-02 data. 
The spatial distributions $\rho({\bf r})$ of SNRs and PWNe are modeled with a smooth function, taken from Ref.~\cite{Green:2015isa} for the former and Ref.~\cite{2004IAUS..218..105L} for the latter source population. 
These functions have been derived on the most updated samples of detected SNRs and pulsars (for PWNe), and corrected for source selection effects. 
Our source distributions are uniformly smooth in all the Galaxy, and do not account for possible single, bright sources in the few kpc around the Earth \cite{Manconi:2016byt,Manconi:2018azw}, or source stochasticity \cite{Manconi:2018azw,Manconi:2020ipm,Mertsch:2018bqd}. 
We also consider the addition of the spiral arms structures for the PWN and SNR spatial distributions using the four-arm structure described in Ref.~\cite{1992ApJS...83..111W} updating the parameters to the ones provided by Ref.~\cite{2006ApJ...643..332F}. 
The implementation of the spiral arms presence in our semi-analytical technique closely follows  Ref.~\cite{DiMauro:2017jpu}, to which we adress for further details.

We anticipate that the main conclusions of the paper remain unchanged when adding a spiral arm pattern in the SNR and PWN spatial distributions. For this reason, our benchmark models are defined without the spiral arm pattern, while we will comment on the effect of its inclusion in specific examples. 

We model the injection spectrum of $e^-$ by SNRs with an energy power law with an exponential cutoff.
We fix the cutoff energy at 20 TeV since there is no evidence of a cutoff in the $e^-$  data  \cite{PhysRevLett.122.101101}. 
Instead, the PWN spectrum is calculated by taking a broken power law around $E_b=500$ GeV, since the $e^+$ AMS-02 data shows a significant softening above a few hundreds of GeV.
We test also values of $E_b$ = 300 and 700 GeV  and we find better fits with $E_b$ = 500 GeV for all the tested models. 
Moreover, a broken power-law energy spectrum is suggested by multiwavelenght observations of PWNe \cite{Bykov:2017xpo}. Both SNRs and PWNe are modeled as burst like events for which all $e^{\pm}$ are injected in the ISM at the time of the supernova explosion.

The propagation of $e^{\pm}$ in the Galactic diffusive halo, of radius $r_{\rm disc}=20$~kpc and vertical half height $L\simeq 1$-$15$~kpc, is calculated through the transport equation:
\begin{equation}
\label{eq:transport}
 \partial_t \psi  - \mathbf{\nabla} \cdot \left\lbrace K(E)  \mathbf{\nabla} \psi \right\rbrace + \partial_E \left\lbrace b(E) \psi \right\rbrace = Q(E, \mathbf{r}, t),
\end{equation}
where $\psi= \psi(E, \mathbf{r}, t)$ is the $e^{\pm}$ number density at energy $E$, Galactic position $\mathbf{r}$ and time $t$, being the flux on Earth $\phi = v/4\pi \psi $.
We fix the Galactocentric distance of Earth to be $r_{\odot}=8.33$~kpc.
This differential equation accounts for the energy losses $b(E)$ due to ICS and synchrotron emission (see next section for more details), diffusion on the irregularities of the Galactic magnetic fields parameterized by the diffusion coefficient $K(E)$, and the source terms $Q(E, \mathbf{r}, t)$.
Other processes usually taken into account for CR nuclei are negligible for the propagation of $e^{-}$ (see, e.g., \cite{Evoli:2016xgn}).
In particular, the effect of convective winds and diffusive reacceleration for the $e^\pm$ transport in the Galaxy has been quantified in Refs~\cite{Delahaye:2008ua,Boudaud:2016jvj}.  
The effect of the aforementioned processes is estimated to be at most 10-20\% at 10~GeV, and quickly decreasing with higher energies for propagation parameters similar to the ones adopted here \cite{Boudaud:2016jvj}.
To test the effect of these processes and other low energy effects on our main conclusions, we produce supplemental results by solving the complete transport equation fully numerically, as described in Appendix~B.

The solution of the propagation equation in Eq.~\ref{eq:transport} for a smooth distribution of sources with density $\rho({\bf r})$ and Galactic SN/pulsar rate $\Gamma_{*}$  is found, according to the semi-analytical model extensively described in \cite{DiMauro:2014iia}, assuming homogeneous energy losses and diffusion:
\begin{eqnarray}
\phi(\mathbf{r}_\odot, E&) &= \frac{v}{4\pi} \frac{\Gamma_{*}}{b(E)} \int dE_s \, Q(E_s) \times \nonumber \\
&&  \times \int d^3{\bf r}_s \,\mathcal{G}_r(\mathbf{r}_\odot, E \leftarrow \mathbf{r}_s, E_s ), \,\rho({\bf r}_s),
 \end{eqnarray}
where $\mathcal{G}_r(\mathbf{r}_\odot, E \leftarrow \mathbf{r}_s, E_s )$ is the Green function that accounts for the probability for $e^{\pm}$ injected at $\mathbf{r}_s$ with energy $E_s$ to reach the Earth with degraded energy $E$.
We fix $\Gamma_{*} = 1$/century both for SNR and PWNe. 
The normalization of the source term for the smooth SNR component will be fitted to the data, and expressed in terms of the total energy released in $e^-$, $W_{\rm SNR}$, in units of erg. As for the smooth distribution of PWNe, we assume an average initial rotation energy $W_0=10^{49}$~erg, following the characteristics of ATNF pulsars  obtained with a typical pulsar decay time of $\tau_0=10$ kyr \cite{Mlyshev:2009twa,Manchester:2004bp}. Data are then fitted by adjusting the efficiency $\eta_{\rm{PWN}}$ of conversion of $W_0$ into $e^\pm$ pairs. 


The normalization and slope of the diffusion coefficient $K(E)=\beta K_0 E^\delta$, as well as the half-height of the diffusive halo, are taken from \cite{Genolini:2015cta} (hereafter {\tt Genolini2015}).
In particular,  $K_0 =0.05$~kpc$^2$/Myr, $\delta=0.445$ and $L=4$~kpc. 
Results will be discussed also using the diffusion parameters in \cite{Evoli:2020ash}, for which a broken power law for the diffusion coefficient is assumed (hereafter {\tt BPLDiffusion}).

\section{Energy losses and interstellar radiation field}
\label{sec:modellosses}

The model of the energy losses $b(E)$ due to ICS and synchrotron emission is particularly important for the interpretation of $e^\pm$ flux detected in the AMS-02 energy range. 
In fact, for the transport of high energetic $e^\pm$ in the Galaxy the energy loss timescale is smaller with respect to the diffusion one \cite{Evoli:2016xgn}. 
We consider losses associated to synchrotron emission on the Galactic magnetic field, and ICS losses, which are demonstrated to dominate over other energy loss mechanism for $e^\pm$ observed at Earth with $E>10$~GeV \cite{2010A&A...524A..51D,Evoli:2016xgn}.

We assume a Galactic magnetic field of $3\mu$G \cite{Sun:2007mx,Evoli:2016xgn}, relevant for computing the synchrotron energy losses.
As for the density of the local ISRF, we use as benchmark the model published in \cite{Vernetto:2016alq} (hereafter {\tt Vernetto2016}).
We account for the ICS losses by numerically performing a double integral, both in the photon $\epsilon$ and  electron energy $E$ (and $\gamma=E/(m_e c^2)$), of the Klein-Nishina collision rate \cite{2010A&A...524A..51D}:
 \begin{equation}
 \frac{dE}{dt} = \frac{12 c \sigma_T E}{(m_e c^2)^2} \int_0^{\infty} d\epsilon \,\epsilon \, n(\epsilon) \, \mathcal{J}(\Gamma), 
 \label{eq:losses}
\end{equation}
where $\Gamma=4\epsilon\gamma/(m_e c^2)$, $\sigma_T$ is the Thomson cross section, $n(\epsilon)$ is the ISRF spectrum and $\mathcal{J}(\Gamma)$ is defined as
  \begin{equation}
\mathcal{J}(\Gamma)= \int^1_0 dq \, q \,\frac{ 2 q \log{q} + (1+2q)(1-q) + \frac{(\Gamma q)^2(1-q)}{2(1+\Gamma q)}}{(1+\Gamma q)^3}
 \label{eq:G}
\end{equation}
with $q=\epsilon / (\Gamma (\gamma m c^2 - \epsilon))$.
Our benchmark results are computed by performing the full numerical integration of the Klein-Nishina loss rate (labeled as {\tt ICS numerical}). We also test the approximated function as in \cite{Evoli:2020ash}, which was originally derived in \cite{2010NJPh...12c3044S}, (labeled as {\tt ICS approx}). 
We extensively comment about possible inaccuracies introduced by analytical approximations of the computation of ICS losses in the Appendix.

For the sake of clarity, our benchmark model of energy losses assumes a magnetic field of $3\mu$G and 
the exact value of the ISRF density reported in Ref.~\cite{Vernetto:2016alq} at any frequency. 
However, to test our results against the effects of the energy losses treatment, we also change the ISRF into the one in Ref.~\cite{2010A&A...524A..51D} (hereafter {\tt Delahaye2010}) and Ref.~\cite{Porter_2006} ({\tt Porter2006}), or the one recently used in \cite{Evoli:2020szd} ({\tt Evoli10/2020}).
The ISRF in Refs.~\cite{2010A&A...524A..51D,Evoli:2020szd} are described using a black body approximation. 
The full model for the ISRF spectrum can be indeed fairly well approximated by a sum of black bodies spectra, each peaked at a characteristic temperature.
In what follows we outline the details of the black body approximations we use to model the ISRF, and we describe the different cases that will be used later. 

\begin{figure*}[t]
  \centering
\includegraphics[width=0.49\textwidth]{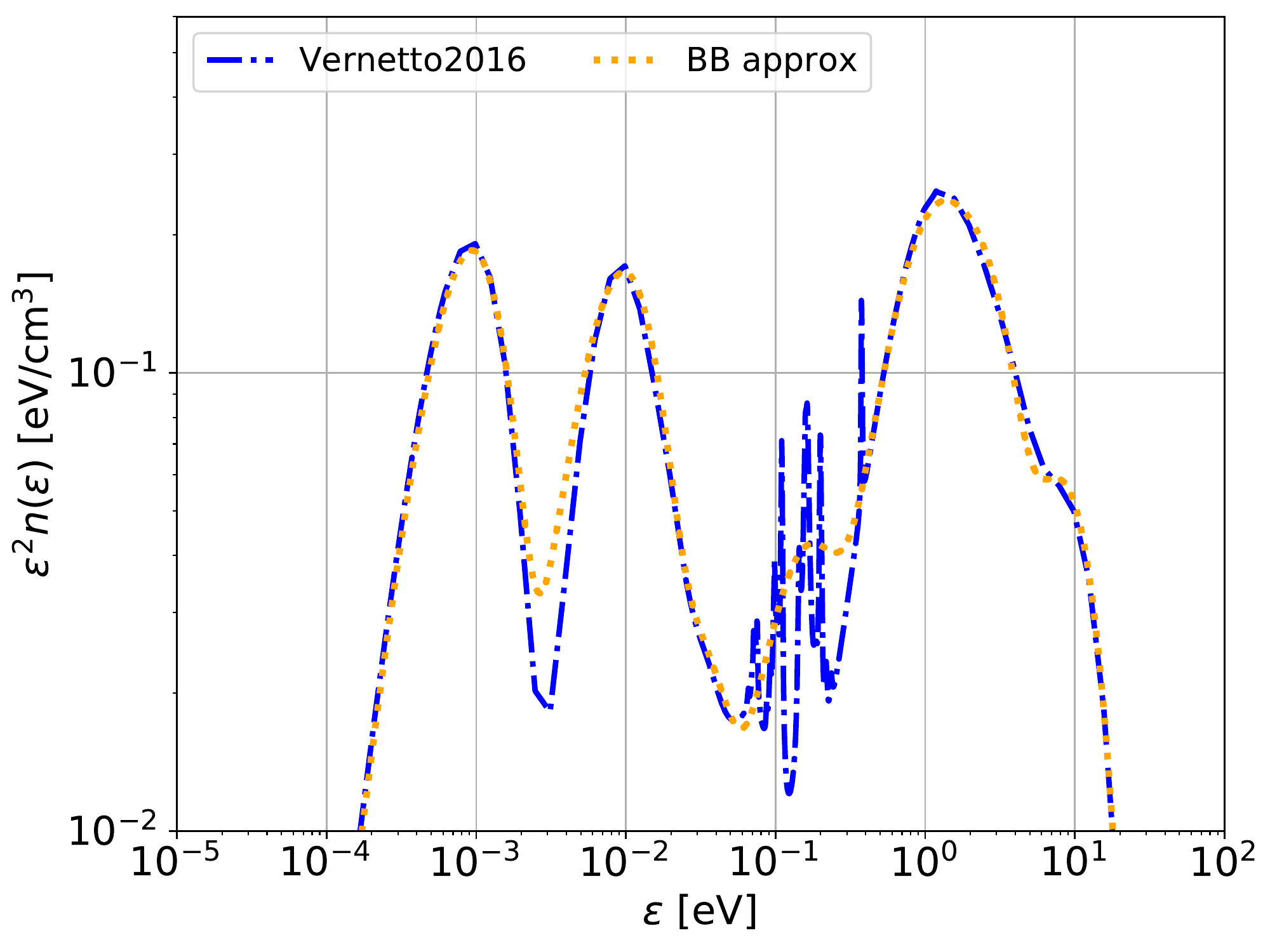}
\includegraphics[width=0.49\textwidth]{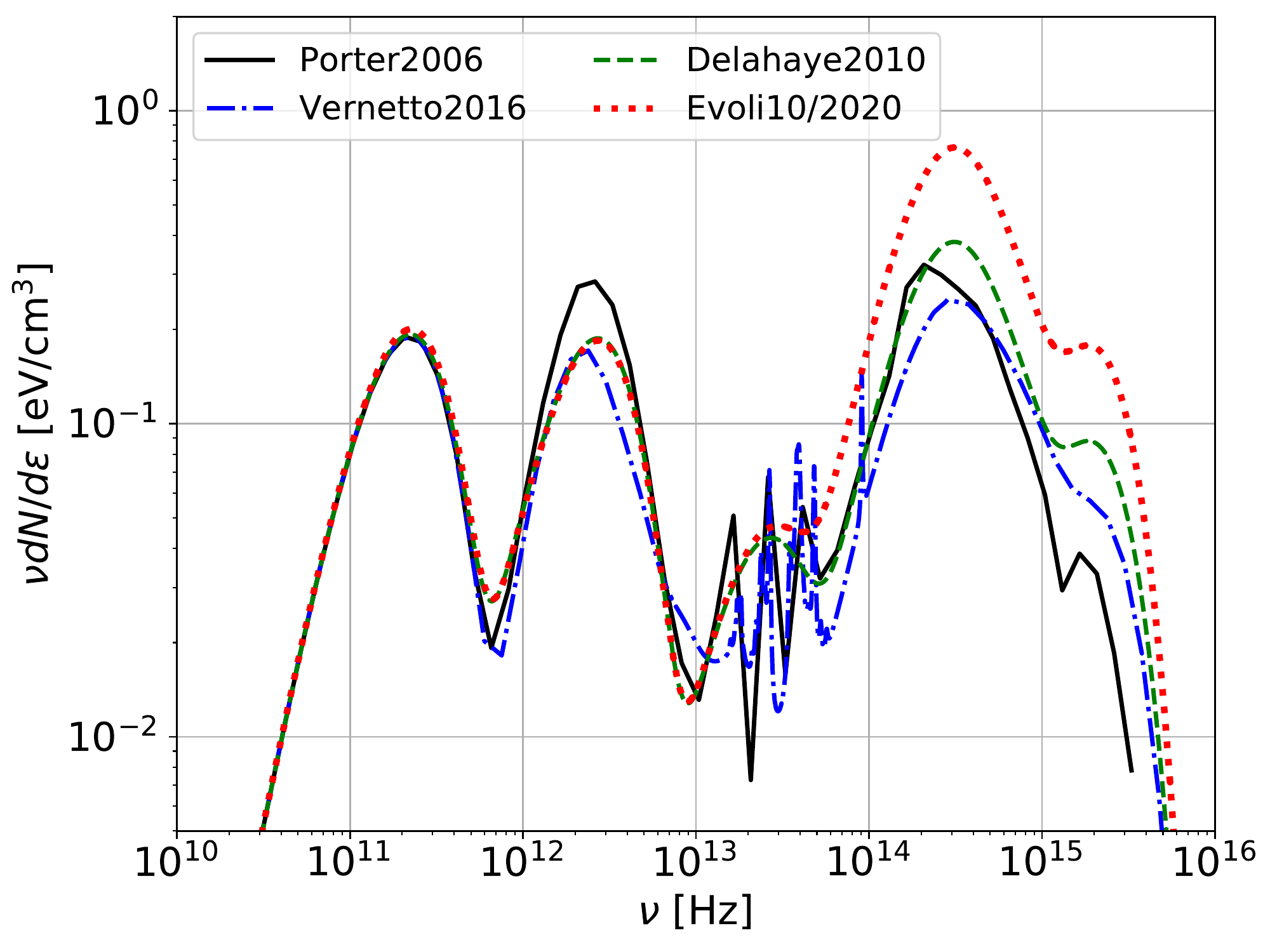}
  \caption{Left panel: Comparison between the spectrum of the ISRF density
  $n(\epsilon)$ published in \cite{Vernetto:2016alq}) (blue dot-dashed line) with the black body approximation introduced in this paper (dotted orange line). 
  Right panel: comparison among the ISRF models {\tt Porter2006} \cite{Porter_2006} (solid black line), 
  {\tt Vernetto2016} \cite{Vernetto:2016alq} (blue dot-dashed line), {\tt Delahaye2010} \cite{2010A&A...524A..51D} (dashed green line) and {\tt Evoli10/2020} \cite{Evoli:2020szd} (dotted red line).}.
  \label{fig:ISRF}
\end{figure*} 

The energy density of the local ISRF provided by Ref.~\cite{Vernetto:2016alq} is illustrated in  Fig.~\ref{fig:ISRF} (blue dot-dashed line). 
The energetics of the ISRF components are as follows. The CMB is at microwave energies, the starlight is between the infrared, visible light and the high-energy tail above 3 eV is at ultraviolet.
Finally, the dust emission is produced by the star light in the optical and ultraviolet that is absorbed and re-emitted at infrared energies.
Following the procedure performed in Ref.~\cite{2010A&A...524A..51D}, we  introduce 7 black body distributions to model the local ISRF spectrum: one for the CMB, three for the dust emission (heated by starlight) and three for the starlight.
The values of the temperature and the energy density of each component are fixed to properly fit the local Galaxy ISRF models published in \cite{Vernetto:2016alq,Porter_2006}.
\begin{table*}
\begin{center}
\begin{tabular}{|c|c|c|c|c|c|c|c|}
\hline
ISRF  &   CMB  &  Dust emission 1  &  Dust emission 2 & Dust emission 3 & Starlight 1 &  Starlight 2  &  Starlight 3 \\ 
\hline
{\tt Vernetto2016}: $T$ [K]  & 2.75  &  28  &  80  &  450  & 2850   & 6150  &  23210  \\
{\tt Vernetto2016}: $u_\gamma$ [eV/cm$^3$]  & 0.25  &  0.22  &  0.049  &  0.032  &  0.18  & 0.24  & 0.079  \\
\hline
\hline
{\tt Porter2006}: $T$ [K]  & 2.75  &  30  &  80  &  450  & 2650   & 5550  &  20210  \\
{\tt Porter2006}: $u_\gamma$ [eV/cm$^3$]  & 0.25  &  0.36  &  0.049  &  0.032  &  0.30  & 0.20  & 0.051  \\
\hline
\hline
\end{tabular}
\caption{This table reports the temperature and energy density  $u_\gamma$ of the different black body components used to fit the ISRF model in Refs. \cite{Vernetto:2016alq} and \cite{Porter_2006}, labeled as {\tt Vernetto2016} and {\tt Porter2006}, respectively.}
\label{tab:ISRF}
\end{center}
\end{table*}
In Tab.~\ref{tab:ISRF} we report the values for the temperature $T$ and the energy density   $u_\gamma$ that we obtain for each of the 7 ISRF components, where $u_\gamma = \int_0^\infty  \epsilon n(\epsilon)d\epsilon$ and  $n(\epsilon)$ is the black body energy distribution. 
The black body energy distributions in the approximated approach are shown in the left panel of Fig.~\ref{fig:ISRF} (orange dotted line) for the {\tt Vernetto2016} model. 
As clearly visible, the black body method provides a very good representation of the ISRF at the peaks of the CMB, dust emission and starlight, and also at the transitions of the three components.
The only part of the ISRF spectrum that the black body approximation is not able to properly capture is between $10^{13}-10^{14}$ Hz, where the full model contains lines in the spectrum associated to the absorption of star light by dust.
The right panel of Fig.~\ref{fig:ISRF} illustrates different modelings of the local ISRF:  
{\tt Porter2006} \cite{Porter_2006} (solid black line), 
  {\tt Vernetto2016} \cite{Vernetto:2016alq} (blue dot-dashed line), {\tt Delahaye2010} \cite{2010A&A...524A..51D} (dashed green line) and {\tt Evoli10/2020} \cite{Evoli:2020szd} (dotted red line).
We note that all the models are very similar across all the frequency interval, apart for {\tt Evoli10/2020} around $10^{14}-10^{16}$~Hz.
This is explained by the choice of \cite{Evoli:2020szd} to multiply the model initially used in \cite{Evoli:2020ash} by a factor $\sim2$ in  this frequency range. 
This factor has been  obtained  by computing  the ratio of the starlight and dust emission on a scale of $\sim5$ kpc around the Sun.
However, the fact that the starlight is twice the dust component averaging over the Galactic halo, does not guarantee that it is sufficient to multiply by a factor of two the local density of the former photon field to obtain the correct ISRF. Moreover, $e^{\pm}$ propagating in the Galaxy above 10 GeV spend most of their time close to the Galactic disk \cite{2010A&A...524A..51D}.
Nevertheless, we explore the consequences of this modification in the local ISRF on the interpretation of the $e^-$ spectrum in the next section. 

\begin{figure}[t]
\centering
\includegraphics[width=0.49\textwidth]{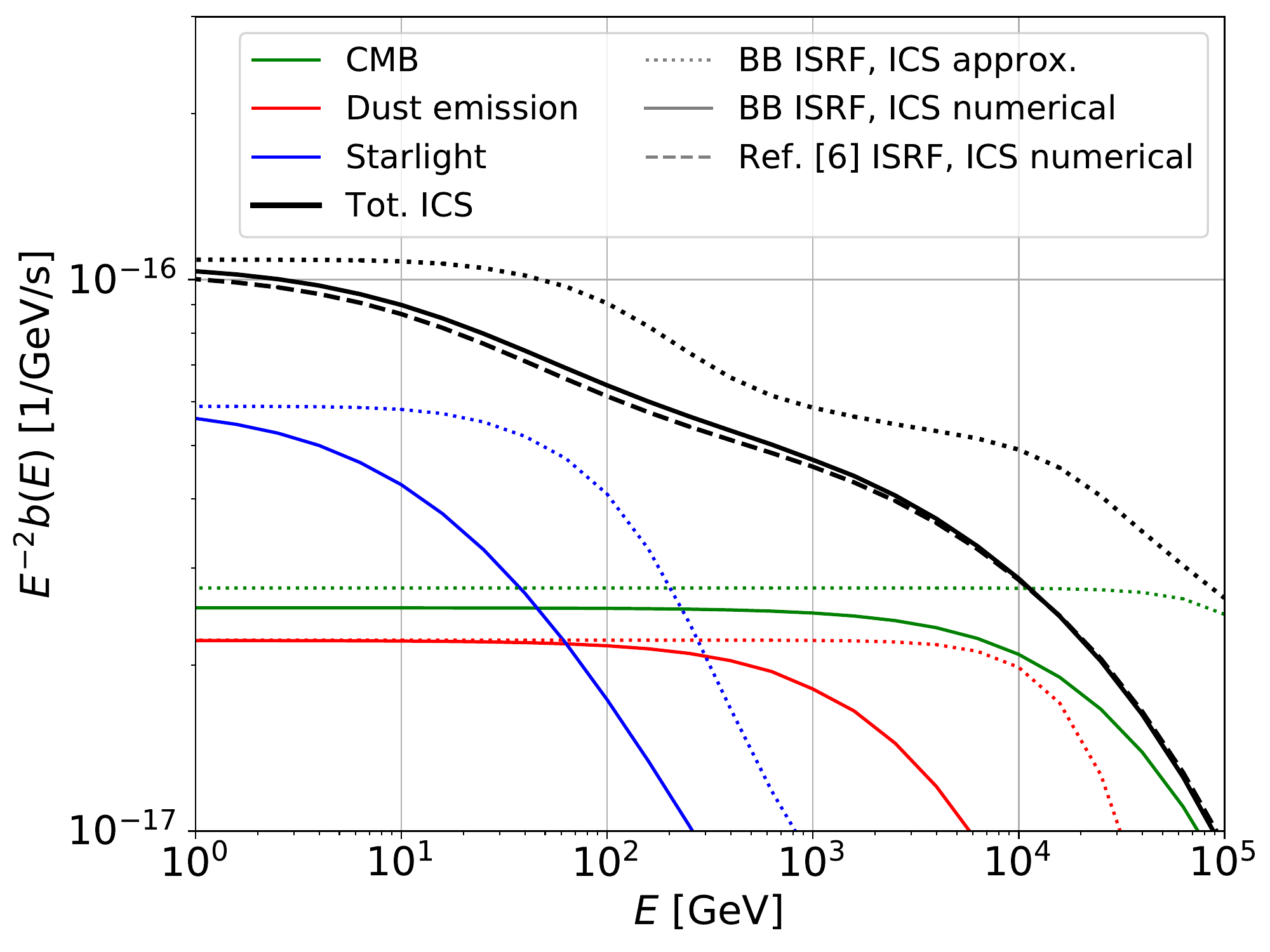}
\caption{Energy loss rate  $b(E)= dE/dt$ (multiplied by $E^2$) for ICS off the ISRF photons composed by CMB (green lines), dust emission (red lines) and starlight (blue lines), for $e^{\pm}$ energy $E$. The total rate is shown with black lines. 
We report three cases: black body approximations of the ISRF and approximated Klein-Nishina calculation as in \cite{2010NJPh...12c3044S} (dotted lines), black body approximations of the ISRF and full numerical Klein-Nishina calculation (solid lines), {\tt Vernetto2016}  ISRF model  and full numerical Klein-Nishina calculation (no approximations, dashed line).}
\label{fig2:losses}
\end{figure}

In Fig.~\ref{fig2:losses} we show the energy loss term $b(E)= dE/dt$ (multiplied by $E^2$) as a function of $e^{\pm}$ energy $E$. 
The energy loss rate has been obtained for each ISRF photon field, both for the approximated Klein-Nishina formalism in Ref.~\cite{2010NJPh...12c3044S} and for the full numerical integration (see Eq.~\ref{eq:losses}). Our reference case, with the {\tt Vernetto2016} ISRF a and the full numerical Klein-Nishina implementation, corresponds to the black solid line.
It is clearly visible that the {\tt ICS approx} cases are significantly different from the  Klein-Nishina exact calculation ones for each photon field, in particular when the Thomson regime does not apply.
We find that only the ICS calculated using the {\tt ICS approx} exhibits a visible change of slope, due to the transition from the $e^{\pm}$ scattering on the starlight to the one on the dust emission and CMB. 
Similar differences in the energy losses calculated with the numerical integration or approximated Klein-Nishina in \cite{2010NJPh...12c3044S} have been recently noted also in \cite{Fang:2020dmi}.
On the other hand, the energy loss rates calculated performing the full numerical integration of the Klein-Nishina rate are smooth, and roughly compatible with a power-law $b(E) \sim E^{1.9}$ from a few GeV to a few TeV.
We report in the Appendix a detailed discussion of the approximated computation shortcomings. 
Finally, we verified that the difference in the ICS energy loss rate by using the complete information for the ISRF density as in {\tt Vernetto2016} and {\tt Porter2006}, or our black body approximation, is at most 4-5$\%$ at low electron energy $E$, and basically negligible for $E > 1$ TeV. This is shown in Fig.~2  for the {\tt Vernetto2016} ISRF model, while very similar results are found also for  {\tt Porter2006}.

\section{Results} 
\label{sec:results}

\subsection{Energy loss rate}

\begin{figure}[t]
  \centering
  \includegraphics[width=0.49\textwidth]{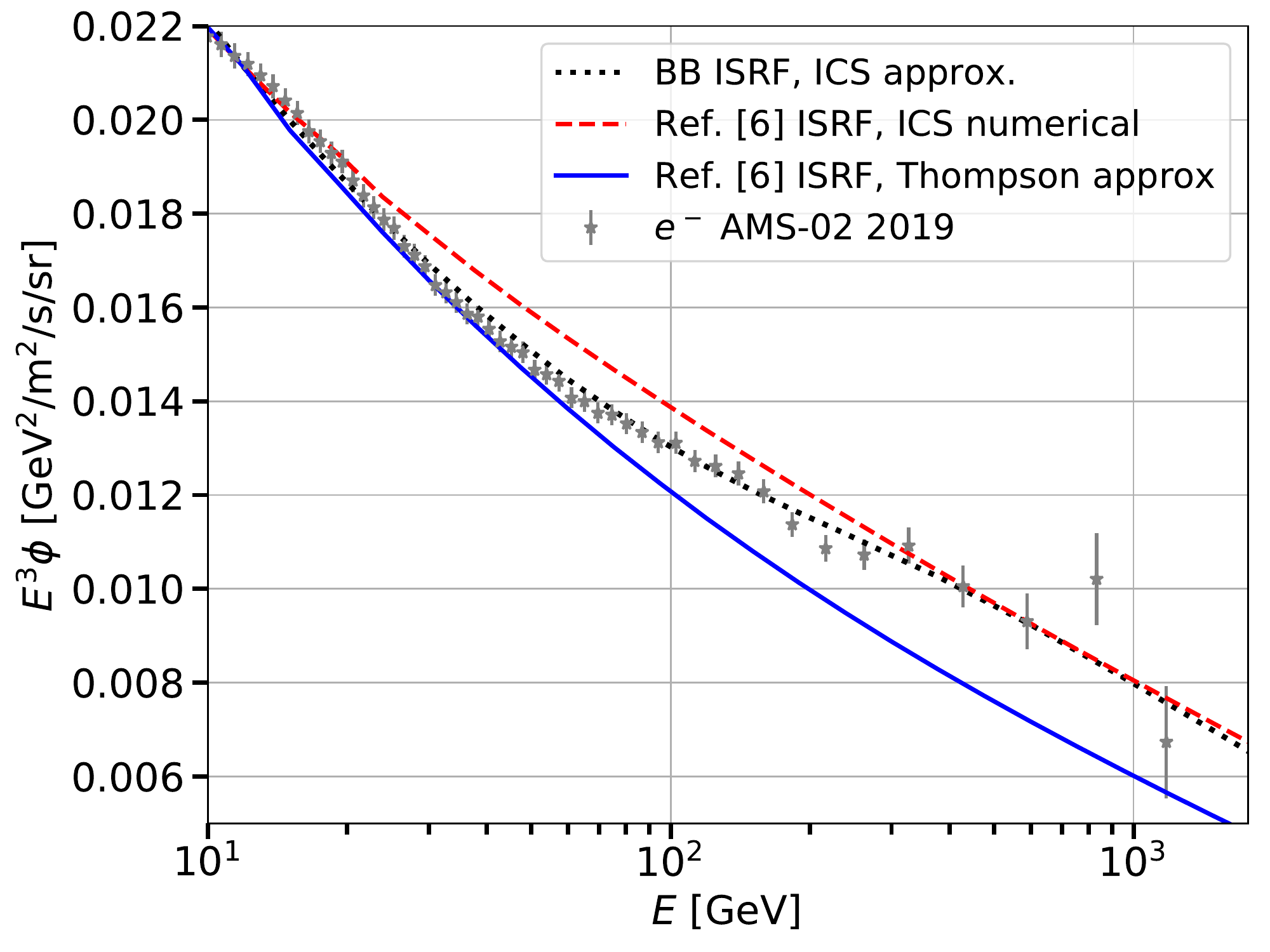}
  \caption{Flux of $e^-$ from a smooth distribution of SNRs calculated 
  for $\gamma_{\rm{SNR}} =2.55$.  We show the same cases for the ISRF and the Klein-Nishina energy loss rate as the ones reported in Fig.~\ref{fig2:losses}, in order to demonstrate the effect of the approximated calculation of the ICS energy losses published in Ref.~\cite{2010NJPh...12c3044S} and implemented in Ref.~\cite{Evoli:2020ash} on the $e^-$ flux.}
  \label{fig3:SNRflux}
\end{figure}

Firstly, we verify the impact of the Klein-Nishina cross section approximated treatment in Ref.~\cite{2010NJPh...12c3044S} with respect to the fully numerical case on the flux of $e^-$ from SNRs.
We implement the $b(E)$ cases reported in Fig.~\ref{fig2:losses} with the Klein-Nishina loss rate ({\tt ICS numerical}) and the approximated treatment ({\tt ICS approx}). The $e^-$ flux computed using the Thomson approximation is also reported for comparison. We fix $\gamma_{\rm{SNR}} = 2.5$ and the 
{\tt Vernetto2016} ISRF density. The result is shown in Fig.~\ref{fig3:SNRflux}.
We normalize the flux of the three cases in such a way that they have the same value at 10 GeV.
The {\tt ICS approx} case provides the same flux values with respect to the {\tt ICS numerical} case 
 at the lowest and highest energies considered: 10 and 2000 GeV. 
Instead, it exhibits a softer spectrum between 10 and 100 GeV, and an hardening at higher energies.
This spectral change makes the {\tt ICS approx} case follow the same trend of the AMS-02 data between 10 and 100~GeV, and is roughly on top of the data above a few hundreds GeV.
On the other hand, the flux predicted with the {\tt ICS numerical} cases
does not show any evident change of shape over the whole energy range, thus suggesting to exclude the option that the break in the $e^-$ AMS-02 data might be due to the ICS energy losses.
We show also the case with ICS losses calculated in the Thomson regime. This case is compatible with the data between 10 and 60 GeV and becomes softer than the other two cases, as well as the data, at higher energies. This is explained by the fact that the energy losses in the Thomson regime are stronger, and thus high-energy $e^-$ lose more energy in the propagation through the Galaxy.
We therefore expect that by fitting the AMS-02 $e^-$ data within Thomson approximation we would need harder indexes with respect to the other two cases shown in Fig.~\ref{fig3:SNRflux}.
We also test the {\tt Porter2006} and {\tt Evoli10/2020} ISRF models, as well as different propagation setups and spectral indexes for the SNR emission, finding similar results.

\begin{figure*}
\centering
\includegraphics[width=0.49\textwidth]{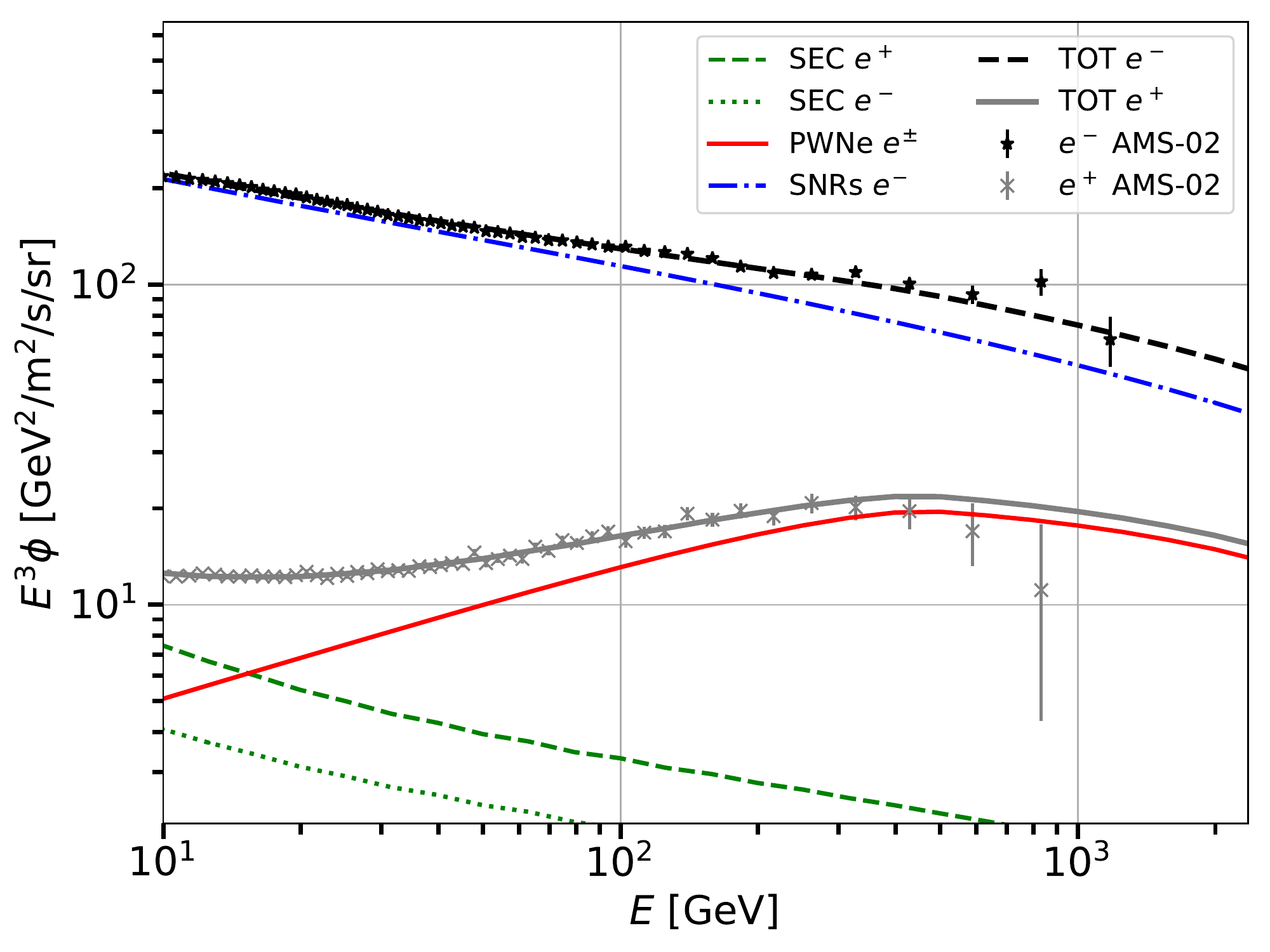}
\includegraphics[width=0.49\textwidth]{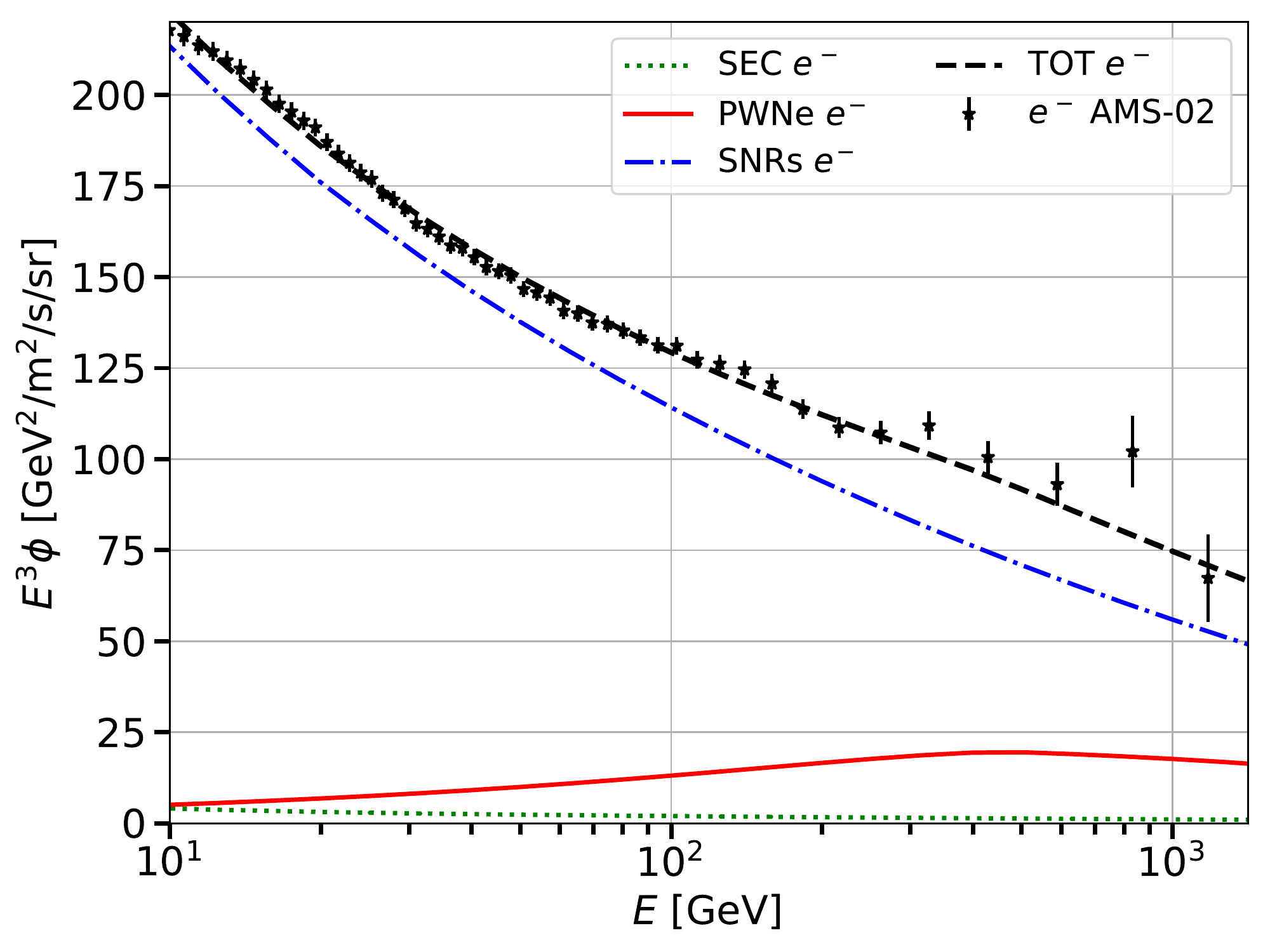}
\caption{Left Panel: Result for the combined fit to $e^{-}$ and $e^{+}$ AMS-02 data (black and grey data points). We show the secondary production of $e^+$ (dashed green line) and $e^{-}$ (dotted green line), $e^\pm$ from PWNe (solid red line), $e^-$ from SNRs (dot-dashed blue line). Right Panel: same as the left panel but zooming in the $e^{-}$ sector.}
\label{fig1:fit}
\end{figure*}

\begin{table*}
\begin{center}
\begin{tabular}{|c|c|c|c|c|c|c|c|c|c|c|c|}
\hline
\% & ISRF  &   Propagation  &  Spiral Arms  &   ICS & $q$ & $\gamma_{SNR}$ & $W_{\rm{SNR}}$ &  $\gamma_{1,2}$ &  $\eta_{\rm{PWN}}$  &  $\tilde{\chi}^2$ &  $\sigma_{PWN}$ \\ 
\hline
  &  &     &    &    &  &  & [$10^{49}$ erg] &   &   &  &  \\ 
\hline
1 & {\tt Vernetto2016}  & {\tt Genolini2015}  &  No  &  numerical  &  1.32  &  2.57  & 1.35 &  1.88/2.31  & 0.009 & 0.92 & 5.8 \\
2 & {\tt Vernetto2016}  & {\tt Genolini2015}  &  Yes &  numerical  &  1.54  &  2.43  & 1.53 &  1.61/2.20  & 0.017 & 1.64 & 8.2 \\
3 & {\tt Vernetto2016}  & {\tt BPLDiffusion}     &  No  &  numerical  &  1.32  &  2.50  & 1.15 &  1.80/2.58  & 0.010 & 0.82 & 4.0 \\
4 & {\tt Delahaye2010}     & {\tt Genolini2015}  &  No  &  numerical  &  1.31  &  2.59  & 1.44 &  1.90/2.27  & 0.009 & 0.95 & 6.1 \\
5 & {\tt Delahaye2010}     & {\tt BPLDiffusion}     &  Yes  & approx   &  1.78  &  2.43  & 2.13  &  1.56/2.80  & 0.018 & 0.71 & 0.2 \\
6 & {\tt Evoli10/2020}     & {\tt BPLDiffusion}     &  Yes  & numerical   &  1.50  &  2.56  & 3.34  &  1.82/2.21  & 0.022 & 0.84 &  3.9 \\
\hline
\hline
7 & {\tt Evoli10/2020}     & {\tt Genolini2015}  &  No  &  numerical  &  1.31  &  2.66  & 2.17 &  1.98/2.39  & 0.011 & 0.89 & 4.0 \\
8 & {\tt Porter2006}    & {\tt Genolini2015}  &  No  &  numerical  &  1.35  &  2.58  & 1.24 &  1.87/2.23  & 0.008 & 1.00 & 6.6 \\
9 & {\tt Vernetto2016}     & {\tt Genolini2015}     &  No  & approx   &  1.41  &  2.56  & 1.38  &  1.84/2.80  & 0.008 & 0.71 & 1.1 \\
10 & {\tt Evoli10/2020}     & {\tt BPLDiffusion}     &  No  & numerical   &  1.42  &  2.60  & 1.95  &  1.88/2.65  & 0.011 & 0.78 & 6.4 \\
\hline
\hline
\end{tabular}
\caption{Summary of the results obtained with the combined fit to $e^+$ and $e^-$ AMS-02 data (see text for details). We show cases where we vary the ISRF model, diffusion parameters, where we include or not the Galactic spiral arms, use the numerical calculation of the Klein-Nishina ICS energy losses or employ the approximation in \cite{2010NJPh...12c3044S}. We list the best-fit values for the secondary renormalization factor $q$, SNR spectral index $\gamma_{SNR}$ and average energy emitted per source $W_{\rm{SNR}}$, PWN source spectral indexes $\gamma_1$ and $\gamma_2$ below and above the break energy, the PWN efficiency $\eta_{\rm{PWN}}$ and the value of the best fit reduced chi-square $\tilde{\chi}^2$. The last column reports the significance for the PWN contribution. }
\label{tab:appendix}
\end{center}
\end{table*}

\begin{table*}
\begin{center}
\begin{tabular}{|c|c|c|c|c|c|c|c|}
\hline
ISRF  &   Propagation  &  Spiral Arms  &   ICS & PWN & $\gamma_{SNR}$ & $W_{\rm{SNR}}$  &  $\chi^2$ \\ 
\hline
  &  &     &    &    &  & [$10^{49}$ erg]  &   \\ 
\hline
{\tt Vernetto2016}  & {\tt Genolini2015}  &  No  &  Thomson & No  &  2.47  & 0.94  & 142 \\
{\tt Vernetto2016}  & {\tt Genolini2015}  &  No  &  numerical & No  &  2.53  & 1.18  & 130 \\
{\tt Vernetto2016}  & {\tt Genolini2015}  &  No  &  numerical & Yes  &  2.57  & 1.35  & 89 \\
\hline
{\tt Evoli102020}  & {\tt BPLDiffusion}  &  No  &  Thomson & No  &  2.39  & 0.85  & 137 \\
{\tt Evoli102020}  & {\tt BPLDiffusion}  &  No  &  numerical & No     & 2.54  & 1.60  & 125  \\
{\tt Evoli102020}  & {\tt BPLDiffusion}  &  No  &  numerical & Yes  &  2.60  & 1.95  & 76 \\
\hline
\hline
\hline
\end{tabular}
\caption{Best-fit parameters of SNRs obtained through a fit to $e^{\pm}$ AMS-02 data with the first and last models of Tab.~\ref{tab:appendix}. We report, for each model, in the first (second) row the case for which we calculate the energy losses in the Thomson approximation (Klein-Nishina, numerical) without accounting for the PWNe contribution to the $e^-$ flux. The third row is for the losses calculated with the Klein-Nishina loss rate and adding also the PWNe $e^-$ flux. The last column represents the $\chi^2$ obtained with the combined fit to $e^{\pm}$ AMS-02 data with 97 degrees of freedom.}
\label{tab:ThompKN}
\end{center}
\end{table*}

\begin{figure*}[t]
  \centering
\includegraphics[width=0.49\textwidth]{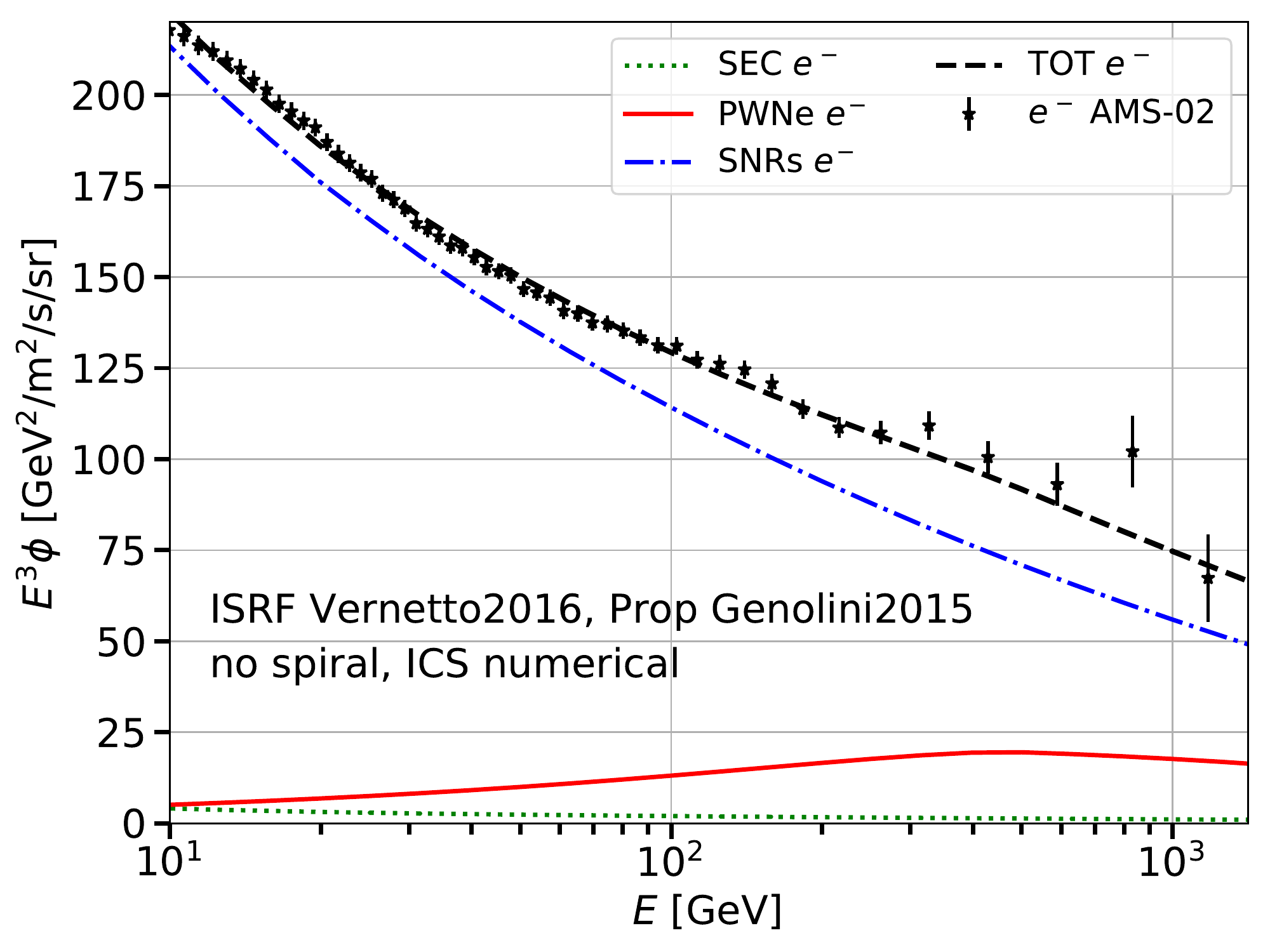}
\includegraphics[width=0.49\textwidth]{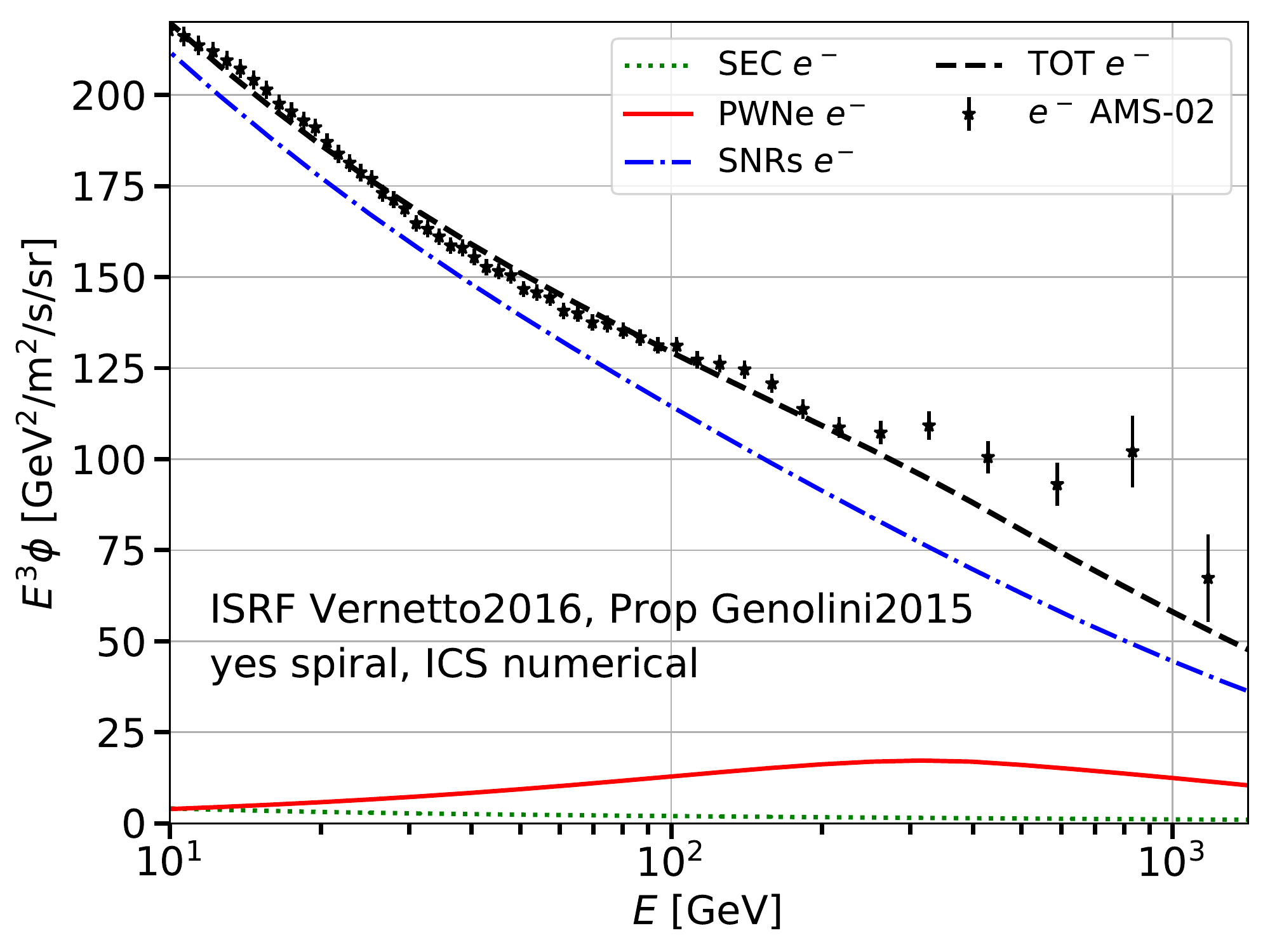}
\includegraphics[width=0.49\textwidth]{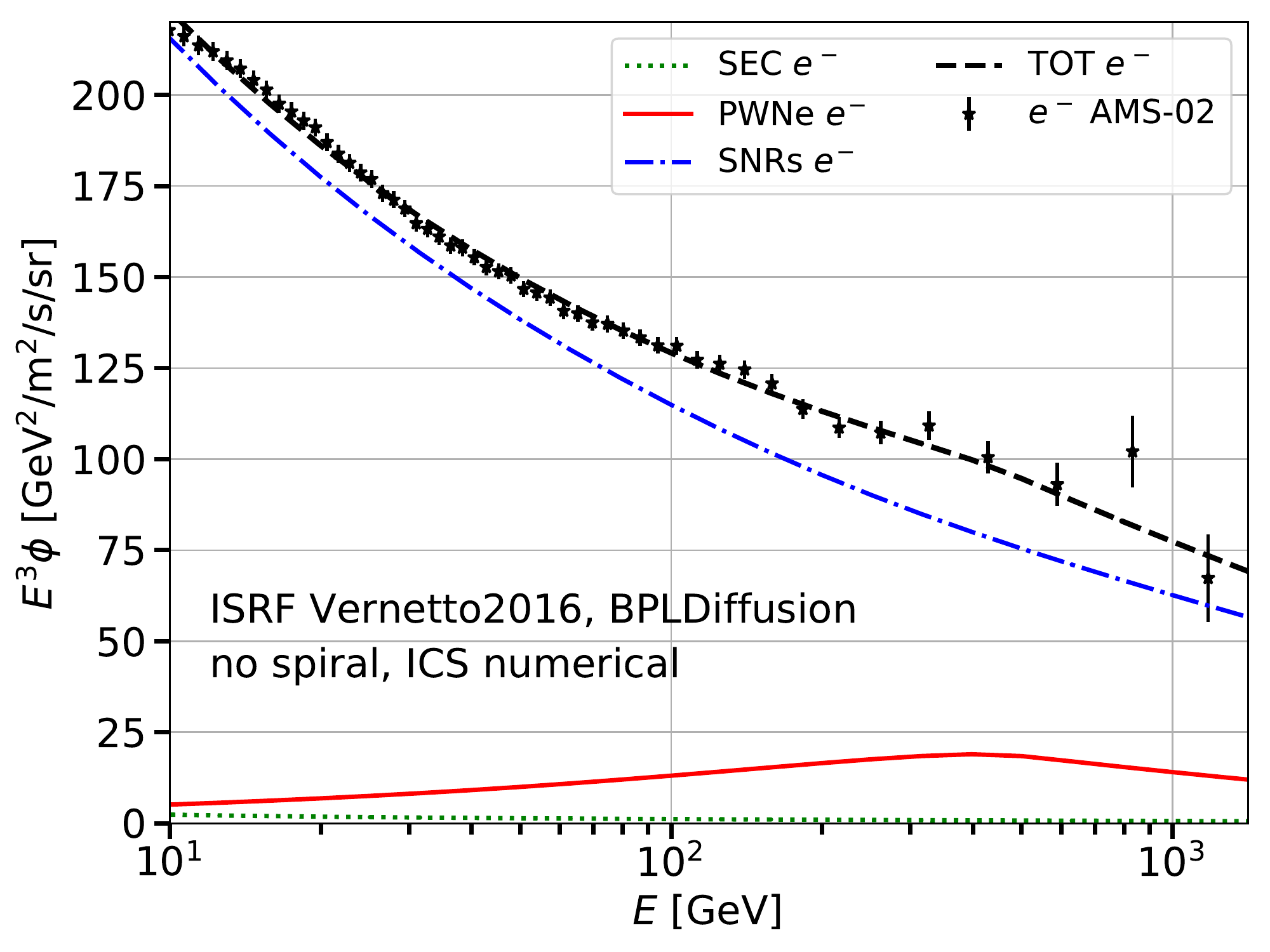}
\includegraphics[width=0.49\textwidth]{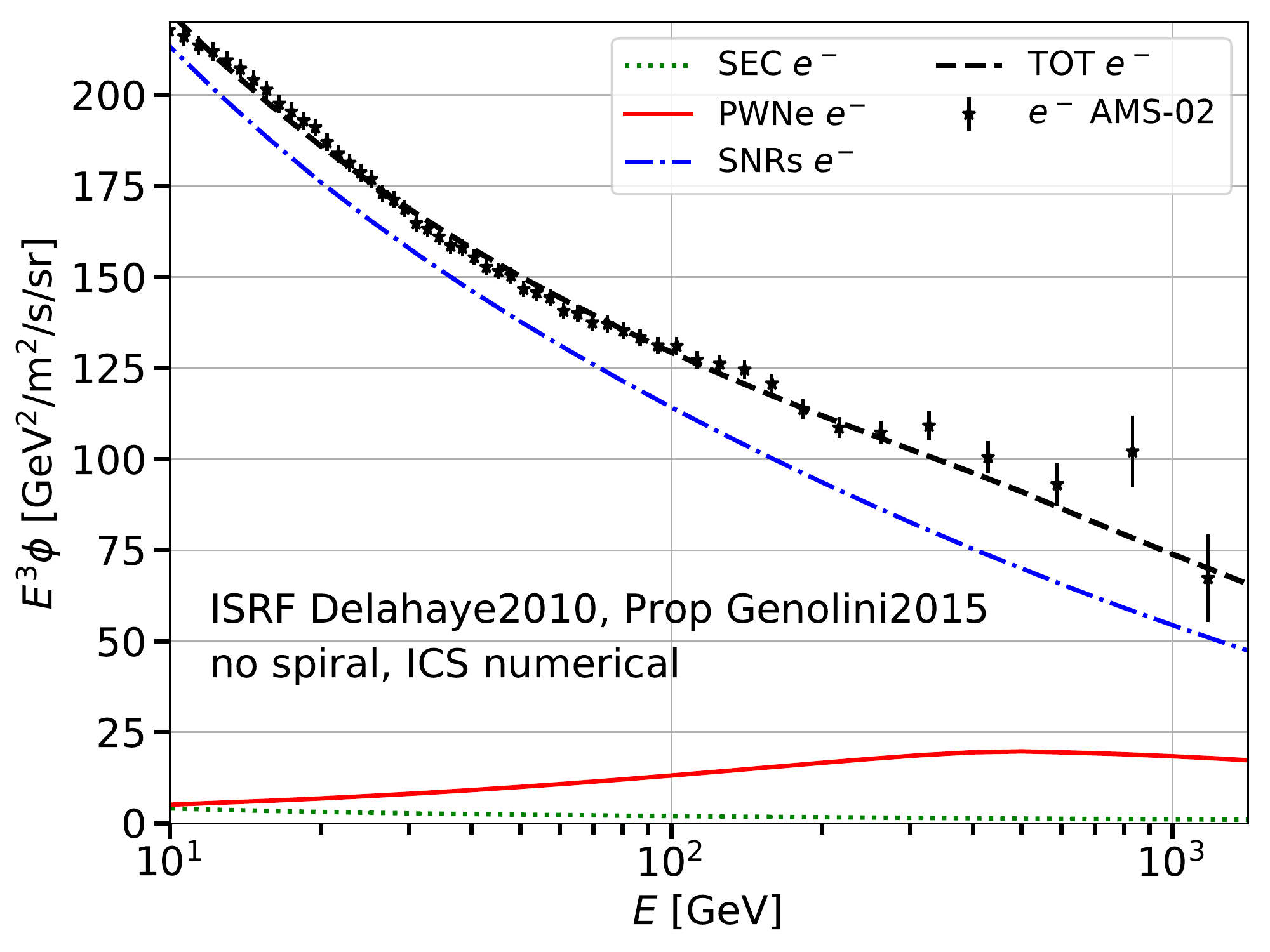}
\includegraphics[width=0.49\textwidth]{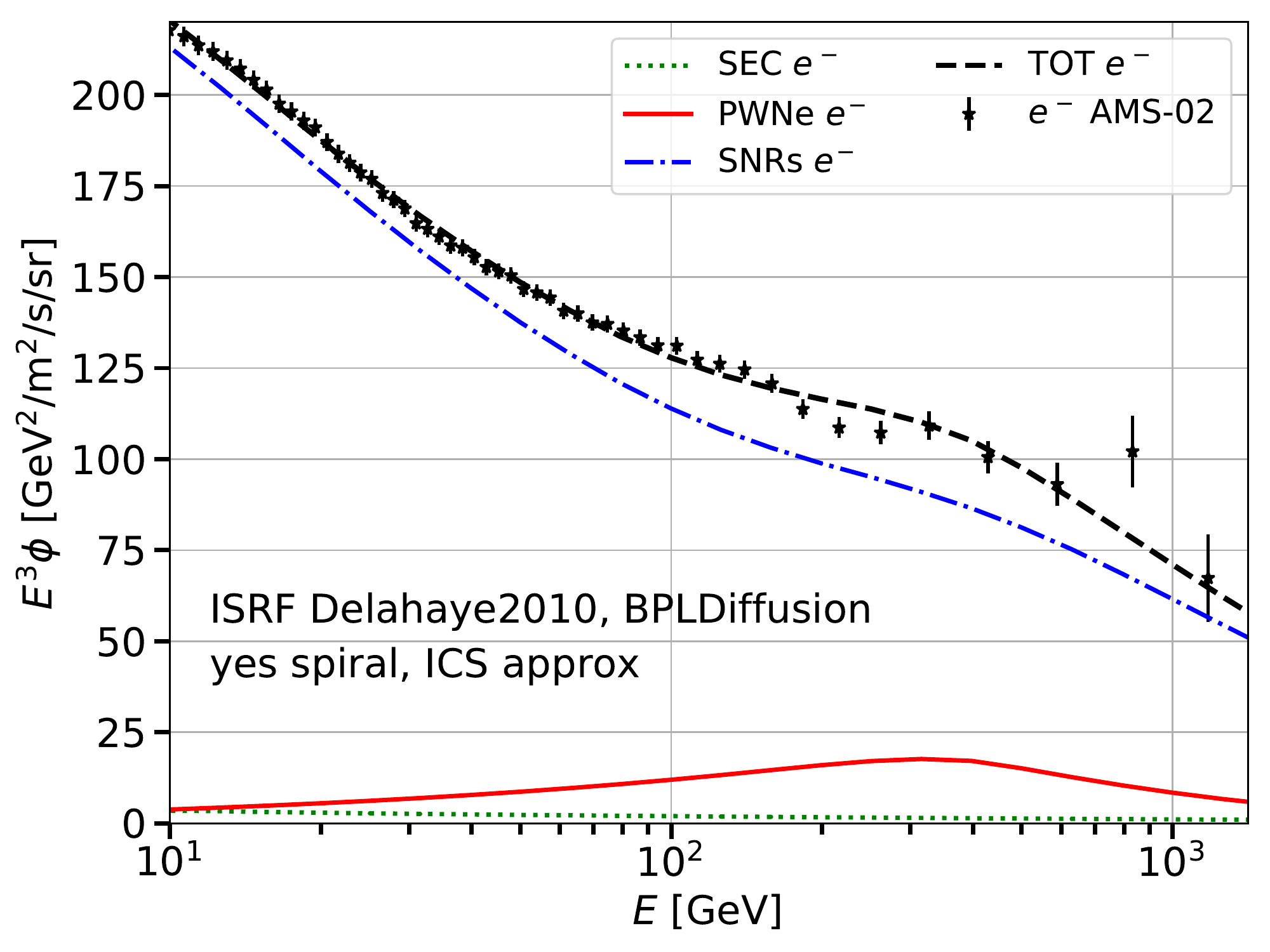}
\includegraphics[width=0.49\textwidth]{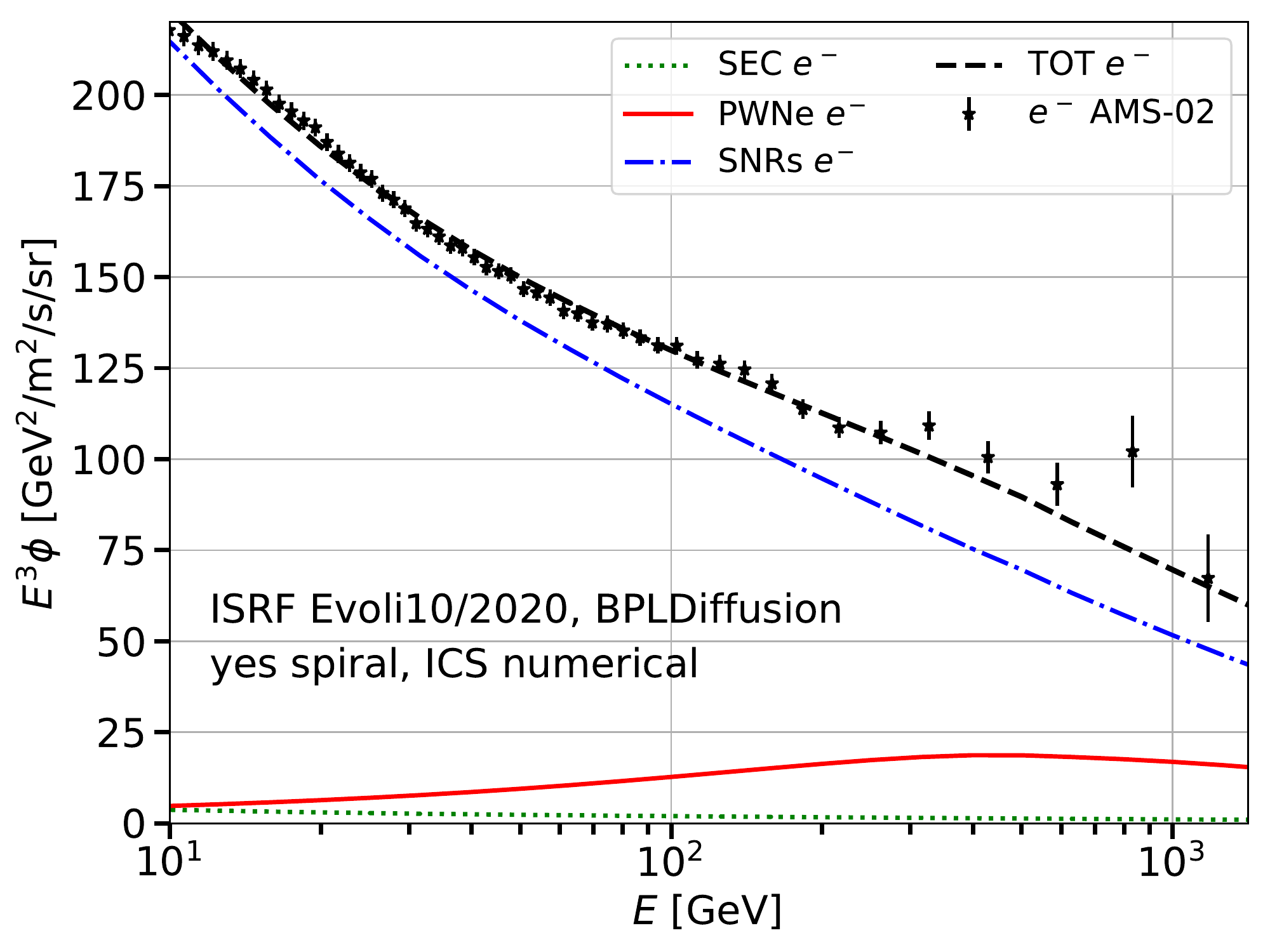}
  \caption{Flux of $e^-$ from SNRs (blue dot-dashed line), PWNe (red solid line) and secondary production (green dotted line) as derived from a combined fit to the $e^{\pm}$ AMS-02 data. We also show the total contribution (black dashed line) and the AMS-02 data (black data points). Each plot refers to one of the first six cases reported in Tab.~\ref{tab:appendix}.}
  \label{fig:fluxes}
\end{figure*}

\subsection{Fit to AMS-02 data}
We then perform a combined fit to the $e^+$ and $e^-$ AMS-02 data above 10 GeV leaving free to vary the normalization of the secondary component $q$, the spectral index $\gamma_{\rm{SNR}}$, the SNR average energy per source $W_{\rm{SNR}}$, the efficiency $\eta_{\rm{PWN}}$ for the conversion of PWN spin-down luminosity into $e^{\pm}$,  and the spectral indexes $\gamma_1$ and $\gamma_2$ below and above the break for the PWN injection spectrum.
We select data above 10 GeV to minimize the effect of the solar modulation that, if not properly taken into account, could generate a bias in the results.
We thus have 6 free parameters in the fit ($q$, $\gamma_{\rm{SNR}}$, $W_{\rm{SNR}}$, $\eta_{\rm{PWN}}$, $\gamma_1$ and $\gamma_2$), and 103 data points.
The fit is performed simultaneously to $e^+$ and $e^-$ data.

We show the results of the fit in Fig.~\ref{fig1:fit} along with the AMS-02 data. 
This is done using the  {\tt Vernetto2016} ISRF and {\tt Genolini2015} propagation parameters.
We find a good agreement with the high-energy part of the $e^{+}$ data with $\gamma_1=1.88$ and $\gamma_2=2.31$. We need an efficiency of about $\eta_{\rm{PWN}}=0.91\%$ that is similar to the value required to explain the $\gamma$-ray halos detected in {\it Fermi}-LAT and HAWC data around the powerful Geminga and Monogem pulsars \cite{DiMauro:2019yvh}.
The best fit for the SNRs provides  $\gamma_{\rm{SNR}}=2.57$ and $W_{\rm{SNR}}=1.4\cdot 10^{49}$ erg, which is compatible with our previous findings on {\it Fermi}-LAT $e^{\pm}$ data \cite{DiMauro:2017jpu}, and similar to the results of Ref.~\cite{Manconi:2018azw}, where we fitted AMS-02 data using also the contributions of single local SNRs. 
The best-fit value for $q$ is 1.32. This implies that we have to renormalize by about $30\%$ the predictions obtained as in \cite{DiMauro:2017jpu}. 
The model reproduces well both the $e^+$ and $e^-$ data in the entire energy range considered.
Indeed, the reduced $\chi^2$ is equal to 0.93.

We also run the same fit procedure using different choices for the ISRF, propagation parameters, the option for spiral arms in the PWN and SNR source distributions, and using the numerical calculation of the ICS energy losses or using the approximated model of Ref.~\cite{2010NJPh...12c3044S}.
In particular we modify each of the ingredients of our benchmark model, that we remind is calculated with the {\tt Vernetto2016} ISRF, {\tt Genolini2015} diffusion coefficient parameterization, 
a smooth spatial distribution of SNRs and PWNe without galactic spiral arms structures, and with the full numerical integrals for the ICS energy losses.  
Specifically, we test our analysis by using the  {\tt Porter2006} and {\tt Delahaye2010} ISRF models. 
We assume the Galactic diffusion parameters {\tt BPLDiffusion}, where the diffusion coefficient is modeled as a broken power law \cite{Evoli:2020ash}.
We also try the addition of Galactic spiral arms structures for the PWN and SNR spatial distributions using the model in \cite{1992ApJS...83..111W}, as already implemented and discussed in Ref.~\cite{DiMauro:2017jpu}. 
This case is named as {\tt Spiral Arms}. 
In a couple of exercises we employ the approximation in \cite{2010NJPh...12c3044S} for the ICS $b(E)$. 
In Tab.~\ref{tab:appendix} we summarize the results of the fits considering these different scenarios.
In the first four cases we have modified our benchmark model by a single ingredient one at a time. The fifth and sixth cases are similar to the model used in Refs.~\cite{Evoli:2020ash,Evoli:2020szd}, while the last four ones are additional combinations of the ISRF, propagation parameters, spiral arms and ICS losses calculation.
We report in Tab.~\ref{tab:appendix} the best-fit values for the secondaries renormalization parameter $q$, the SNR spectral indexes $\gamma_{\rm SNR}$, 
the average energy emitted per source $W_{\rm{SNR}}$, the PWN spectral indexes $\gamma_1$ and $\gamma_2$ below and above the break energy, the PWN conversion efficiency $\eta_{\rm{PWN}}$ and the combined $e^-$ and $e^+$ reduced chi-squared $\tilde\chi^2$ for the best fits.

We find that all the tested cases with spiral arms produce harder SNR and PWN spectral indexes.
For all the tested cases we find a renormalization of the secondary production between $30-50\%$, except for the ICS model used in Ref.~\cite{Evoli:2020ash} with the approximated formula in Ref.~\cite{2010NJPh...12c3044S} and  {\tt Delahaye2010} ISRF, for which we find a factor higher by almost $80\%$. 
Changing the ISRF model provides similar values for the free parameters and the goodness of fit. The  {\tt Vernetto2016} ISRF gives slightly better fits with respect to the other two ones.
Assuming the {\tt BPLDiffusion} propagation model we find slightly better fits, and with very similar best-fit parameters, with respect to the {\tt Genolini2015} scenario. 
Depending on the spectral index of the SNR population, also the {\tt BPLDiffusion} introduces a mild change of slope in the propagated spectrum at high energies.
The two cases that gives the lowest $\tilde{\chi}^2$ are obtained using the approximated computation of ICS energy losses  \cite{2010NJPh...12c3044S}. 
However, we remind that this computation poorly reproduces the transition between the Thomson regime and the Klein-Nishina formalism as detailed in the Appendix. Moreover, the $\tilde{\chi}^2$ values obtained for the different cases are very similar, and none is significantly better than the other ones from a statistical point of view. Finally, we remind here that this $\tilde{\chi}^2$ is obtained with a combined fit to $e^+$ and $e^-$ data and that the focus of this paper is on the $e^-$ flux.

In Fig.~\ref{fig:fluxes} we show the result on the $e^-$ flux at Earth for the first six cases tested in this Section and summarized in Tab.~\ref{tab:appendix}.
When we only modify the ISRF model or propagation parameters with respect to the benchmark model, we obtain very similar contributions from the SNR and PWN fluxes.
Instead, if we use the ICS energy losses approximation as in Ref.~\cite{2010NJPh...12c3044S}, and then implemented in Ref.~\cite{Evoli:2020ash}, we find a change of trend in the SNR flux at around 100~GeV, similar to what found in Ref.~\cite{Evoli:2020ash}. In particular, since the SNR flux for this model shows an hardening with increasing energy, the PWN contribution is forced to be slightly lower than in the other cases, and the resulting fit is better.
However, as demonstrated in this paper (see the Appendix), this model is based on a poor approximated calculation of the ICS energy losses, which we have shown to poorly reproduce the transition between the Thomson regime and Klein-Nishina formalism.
The benchmark model, in which energy losses are computed using a fully numerical approach, fits very well the data.

While it is clear that the PWN flux is necessary to fit the $e^+$ excess at energies larger than 10~GeV, the statistical significance for this component in contributing to the $e^-$ data, and in particular the observed change of slope at 40-50~GeV, has never been investigated in the literature.
We thus derive the significance for the PWN contribution by calculating the chi-square with the model that contains secondary production and flux from SNRs ($\chi^2_{\rm{noPWNe}}$) and the one with the addition of PWNe ($\chi^2_{\rm{yesPWNe}}$), fitting the AMS-02 $e^{\pm}$ flux data above 10 GeV.
Then, we find the difference of the $\chi^2$ of the two models $\Delta \chi_{\rm{PWNe}}^2 = \chi^2_{\rm{noPWNe}}-\chi^2_{\rm{yesPWNe}}$, that are nested since one can move from the case with PWNe to the one without them by simply fixing $\eta_{\rm PWN} = 0$. 
We convert then the value of $\Delta \chi_{\rm{PWNe}}^2$ to the significance for the PWN contribution according to the additional degrees of freedom of the model with PWNe ($\eta_{\rm PWN}, \gamma_1, \gamma_2$). 
For our benchmark model (first row of Tab.~\ref{tab:appendix}) we obtain $\chi^2_{\rm{noPWNe}}=130$ and $\chi^2_{\rm{yesPWNe}}=89$. The case with the PWN contribution has 97 degrees of freedom. Therefore, $\Delta \chi_{\rm{PWNe}}^2=41$ and the significance for the PWN contribution is $5.8\sigma$.
For all the other cases reported in Tab.~\ref{tab:appendix}, the value of $\Delta \chi_{\rm{PWNe}}^2$ ranges between 22 and 75.
Consequently, we obtain a significance for the PWNe contribution between 4.0 and 8.2$\sigma$.
The models with the lowest significance of the PWN flux corresponds to the ones for which the ICS losses are calculated using the approximations in  Ref.~\cite{2010NJPh...12c3044S}. We test this using both the {\tt Vernetto2016} and {\tt Evoli10/2020} ISRF and {\tt Genolini2015} and {\tt BPLDiffusion} propagation models, finding similar result. The main reason for finding low significances in these cases is that, as already demonstrated, the {\tt ICS approx} model produces a steepening of the flux above 40-50 GeV in the SNR flux and can explain the data without a significant contribution of PWNe (see Fig.~\ref{fig3:SNRflux}).

In Refs.~\cite{Evoli:2020ash,Evoli:2020szd} the authors report that the break in the AMS-02 $e^-$ flux is due to the transition between the Thomson regime and the Klein-Nishina formalism of the ICS losses on the starlight ISRF.
As we have demonstrated, the results in Ref.~\cite{Evoli:2020ash} are likely driven by the approximated calculation for the Klein-Nishina loss rate, while in Ref.~\cite{Evoli:2020szd} no statistical probe is provided to further corroborate this interpretation. 
In order to quantitatively test this hypothesis we perform a fit to the $e^{\pm}$ data by using the source fluxes calculated with the Thomson approximation and no PWN contribution to the $e^-$ flux data or the full numerical Klein-Nishina ICS loss rate without and with the PWN flux, and compare the $\chi^2$ values.
In fact, the two cases with Thomson or Klein-Nishina ICS losses are not nested and thus we can not convert the $\Delta \chi^2$ between them into a significance.
We run this test to verify if the largest improvement in the $\chi^2$ of the fit to $e^{\pm}$ data is due to the transition between the Thomson approximation and the Klein-Nishina ICS formalism, or by the addition of the PWNe flux to the $e^-$ data.
We show the results in Tab.~\ref{tab:ThompKN} for our benchmark model as well as using the {\tt Evoli10/2020} ISRF and the {\tt BPLDiffusion} propagation setup which is similar to the model used in Ref.~\cite{Evoli:2020szd}.
We find for both models a relatively small improvement in the goodness of fit between the case of Thomson and the Klein-Nishina ICS losses, the increase in the $\chi^2$ being 12. Instead, the $\chi^2$ improves significantly, by a value of 41 and 37, when adding the PWN flux in the two models tested in Tab.~\ref{tab:ThompKN}. 
The value of $\Delta \chi_{\rm{PWNe}}^2$ varies between 22 and 75 considering all the cases reported in Tab.~\ref{tab:appendix}.
The $\chi^2$ values reported so far are for the combined fit to $e^{\pm}$. The portion of $\chi^2$ relative to the fit to $e^{-}$ data only is, for our benchmark case, equal to 106 with the Thomson ICS losses and no PWNe, 94 for the Klein-Nishina losses and no PWNe and 52 for the Klein-Nishina losses and with PWN flux. Therefore, the fit on $e^-$ data points, that are 52, improves significantly when adding PWNe into the model, while it changes mildly when calculating the losses with the Klein-Nishina formalism with respect to the Thompson approximation.

\subsection{Low-energy effects}
We perform further analysis to check a possible influence of our simplified, semi-analytical propagation model on our main conclusions. For this pourpose, the transport equation taking into account convective winds, diffusive reacceleration and other low-energy effects is solved numerically. For the fit described in what follows, all the  $e^\pm$ fluxes at Earth for the different components (secondaries,  SNR and PWN) have been obtained within the GALPROP code as described in Appendix B.

We first run a fit to the $e^+$ and $e^-$ AMS-02 data at $E>10$~GeV using the flux from secondaries, SNRs and PWNe as computed with GALPROP in the BASE +inj+va propagation of Ref.~\cite{Korsmeier:2021brc}, in which both convection and reacceleration are included. 
In this case, we find a good fit to the AMS-02 data ($\chi^2=80$), while the significance for the PWN contribution is $6.5\sigma$. The spectral index for the SNRs is $\gamma_{SNR} = 2.62$ while the index below and above the break for PWNe are $\gamma_1 =1.87$ and $\gamma_2 = 2.68$.  We thus confirm the main results of the paper for the fit to AMS-02 data at $E>10$~GeV  also considering a more detailed modeling for the propagation of $e^\pm$ in the Galaxy.

We then make a fit including also low-energy AMS-02 data at $E>5$~GeV. In this case, it has been shown that in order to fit the low-energy data a break in the injection spectrum of SNR at few GeV is required, see e.g.~\cite{LI2015267}. For this simple check, we fix the position of this break at 7~GeV with an index before the break fixed at 1.6, as done in \cite{Fornieri_2020}. 
The goodness of fit worsens to $\chi^2=157$ but the significance for the PWN contribution increases to about $9\sigma$. 
As noted by \cite{LI2015267}, when including low-energy data, the addition of another break in the injection spectrum at around 50 GeV improves significantly the fit to the $e^-$ data, see also discussion in~\cite{DiMauro:2017jpu}. 

%
%
We therefore test the significance of the PWN contribution by fitting data at $E>5$~GeV and adding a break in the injection spectrum of SNR at 50 GeV. 
This is done by performing two separate fits: in the first one we include the PWN contribution both in the $e^-$ and in the $e^+$ data, together with the additional break in the SNR injection spectrum. In the second one, we artificially turn off the PWN contribution to the $e^-$ flux only, while keeping the break in the SNR injection spectrum.
In the first case, by adding a break in the injection spectrum of SNR at 50 GeV and fitting the data with a broken power law spectrum with two free indexes below and above the break, the fit finds  $\chi^2\approx 94$. 
The slopes in the SNR injection spectrum are 2.76 and 2.62 below and above the break. 
In the second case, when performing a fit to $e^\pm$ data turning off the PWN contribution to $e^-$, we find a similar $\chi^2$ as in the previous case and a slightly higher change of slope in the SNR injection with slopes below and above the break of 2.74 and 2.55.  
Since the $\chi^2$ in the case with ad without the PWN contribution is almost the same, the significance of their contribution is almost zero.
We note that the interpretation of this break is unclear, as it is not predicted by the most recent studies on the $e^-$ spectrum released by SNRs (see e.g. \cite{Morlino:2021rzv}),  and it requires to introduce at least two new free parameters. In addition, we remind that a PWN component is always needed to explain the $e^+$ data alone, which brings a physical explanation also to an $e^-$ component.

\medskip 

To summarize, following the results of our statistical tests, we demonstrated that the break in the $e^-$ flux data is naturally explained by the interplay of SNRs and PWNe contribution below and above the observed energy of about 40 GeV, where it has been detected by AMS-02.
In fact, we can look into the $e^-$ data and the best fit model contributions in the right panel of Fig.~\ref{fig1:fit} by considering a linear scale that highlights the change of slope in the data and in the model.
In particular, we observe that the SNR contribution follows the data for energies between 10 to about 50 GeV. Above this energy, where the break has been detected by AMS-02, the contribution of SNRs decreases while the  PWNe one increases. 
The SNRs, PWNe and the secondary production contribute about $96/2/2\%$, $92/7/1\%$, $78/21/1\%$ of the total $e^-$ flux at 10/50/400 GeV, respectively. The PWN to SNR flux relative contribution thus increases by a factor 13 from 10 GeV to 400 GeV.

\section{Conclusions} 
\label{sec:conclusions}
In this paper we have demonstrated that the AMS-02 $e^-$ and $e^+$ flux data can be properly explained with the production of CR leptons from SNRs, PWNe and secondary production. Specifically, $e^+$ above 10 GeV are mostly explained with PWNe with a power-law injection spectrum broken at about 500 GeV, and a change of slope below and above the break of about $\Delta \gamma= 0.5$.
SNRs explain most of the $e^-$ flux. Their contribution decreases with energy from 96$\%$ at 10 GeV to 78$\%$ at 500 GeV, while PWNe provide an increasing contribution reaching a maximal 21$\%$ at 500 GeV. 
For the first time, we estimated the significance of the PWN contribution to the $e^{-}$ flux, that varies within $4-8\sigma$, considering different models for the ISRF, source distribution in the Galaxy and propagation parameters within a semi-analytical diffusion model.
We also provided a statistical test to probe the hypothesis that the break at 40 GeV detected in AMS-02 $e^{-}$ data is due to the transition of ICS energy losses between the Thomson regime and the Klein-Nishina formalism on the starlight component. We quantitatively assess  that the improvement in the fit by using the Klein-Nishina loss rate with respect to the Thomson approximation is much smaller than the one obtained with the addition of the PWNe flux in the model.
The stability our results against low-energy effects, such as convection and reacceleration, is checked  by solving numerically the transport equation, as well as adding possible breaks in the injection spectrum of SNRs. \\
We have also implemented an effective model for the $e^-$  flux, where the SNR injection spectrum follows a power-law with a break at around 50 GeV. We find that the data are fitted in this case equally well as in the case of no break in the SNR injection spectrum and a source of $e^-$ from PWN. In the broken power-law scenario, the significance of the PWN contribution is almost zero. However, the physical motivation of this break at 50 GeV is unclear. Instead, the contribution from PWNe provides a more physical and natural explanation to the data since they have to contribute to the observed $e^+$ flux.
We thus conclude that the break measured by AMS-02 in the $e^-$ cosmic flux at $E\sim 40$~GeV is very likely due to the interplay between the contribution of SNRs and PWNe.

{\it Acknowledgements. -}
The work of FD has been supported by the "Departments of Excellence 2018 - 2022" Grant awarded by
the Italian Ministry of Education, University and Research (MIUR) (L.~232/2016).
MDM research is supported by Fellini - Fellowship for Innovation at INFN, funded by the European Union’s Horizon 2020 research programme under the Marie Skłodowska-Curie Cofund Action, grant agreement no.~754496.

\bibliography{paper}

\appendix

\section{Approximated calculation of the inverse Compton energy losses}
\label{sec:appendix}

The ICS accounts for the interaction between CR $e^{\pm}$ and  the photons of the ISRF.
Since the CMB, dust emission and starlight ISRF photons have energies at most of 1-10 eV, during these scatterings CR $e^{\pm}$ lose a part of their energy, while the photons are upscattered to higher energies, typically between the X-ray and $\gamma$-ray bands.
In the Thomson regime, that is valid for $E \epsilon \ll m_e^2 c^4$ ($m_e$ is the electron mass), the typical energy of the scattered photons for ICS is $4\gamma^2 \epsilon$ for an head-on collision.

For large energies, the Thomson regime is no longer valid. The complete calculation of the ICS energy losses, requires a double integration of the Klein-Nishina cross section multiplied for the ISRF energy density $n(\epsilon)$ \cite{2010A&A...524A..51D}:
 \begin{equation}
 \frac{dE}{dt} = \frac{12 c \sigma_T E}{(m_e c^2)^2} \int_0^{\infty} d\epsilon \,\epsilon \, n(\epsilon) \, \mathcal{J}(\Gamma), 
 \label{eq:losses}
\end{equation}
where $\sigma_T$ is the total Thomson cross section, $\Gamma=4\epsilon\gamma/(m_e c^2)$ and $\mathcal{J}(\Gamma)$ is defined as
  \begin{equation}
\mathcal{J}(\Gamma)= \int^1_0 dq \, q \,\frac{ 2 q \log{q} + (1+2q)(1-q) + \frac{(\Gamma q)^2(1-q)}{2(1+\Gamma q)}}{(1+\Gamma q)^3}
 \label{eq:G}
\end{equation}
with $q=\epsilon / (\Gamma (\gamma m c^2 - \epsilon))$.

Ref.~\cite{2010NJPh...12c3044S} introduced an approximated expression for Eq.~\ref{eq:losses}, which allows a fully analytical solution for the energy loss rate:
 \begin{equation}
 \frac{dE}{dt} = \frac{4 \sigma_T c  E^2}{3(m_e c^2)^2} u_{\gamma} \times \mathcal{F}_{KN}(E), 
 \label{eq:lossessch}
\end{equation}
where $\mathcal{F}_{KN}(E)$ is expressed as:
 \begin{equation}
 \mathcal{F}_{KN}(E) = \frac{\frac{45}{64\pi^2} (\frac{m_ec^2}{ K_B T})^2}{\frac{45}{64\pi^2} (\frac{m_ec^2}{ K_B T})^2 + (\frac{E}{m_ec^2})^2}.
 \label{eq:FKN}
\end{equation}
Digging into the approximated ICS losses formulae presented in Ref.~\cite{2010NJPh...12c3044S} and used by Ref.~\cite{Evoli:2020ash}, we remind that two main simplifications were performed in order to obtain a fully analytical solution of the ICS energy losses in all the energy range.

\begin{figure*}[t]
  \centering
\includegraphics[width=0.49\textwidth]{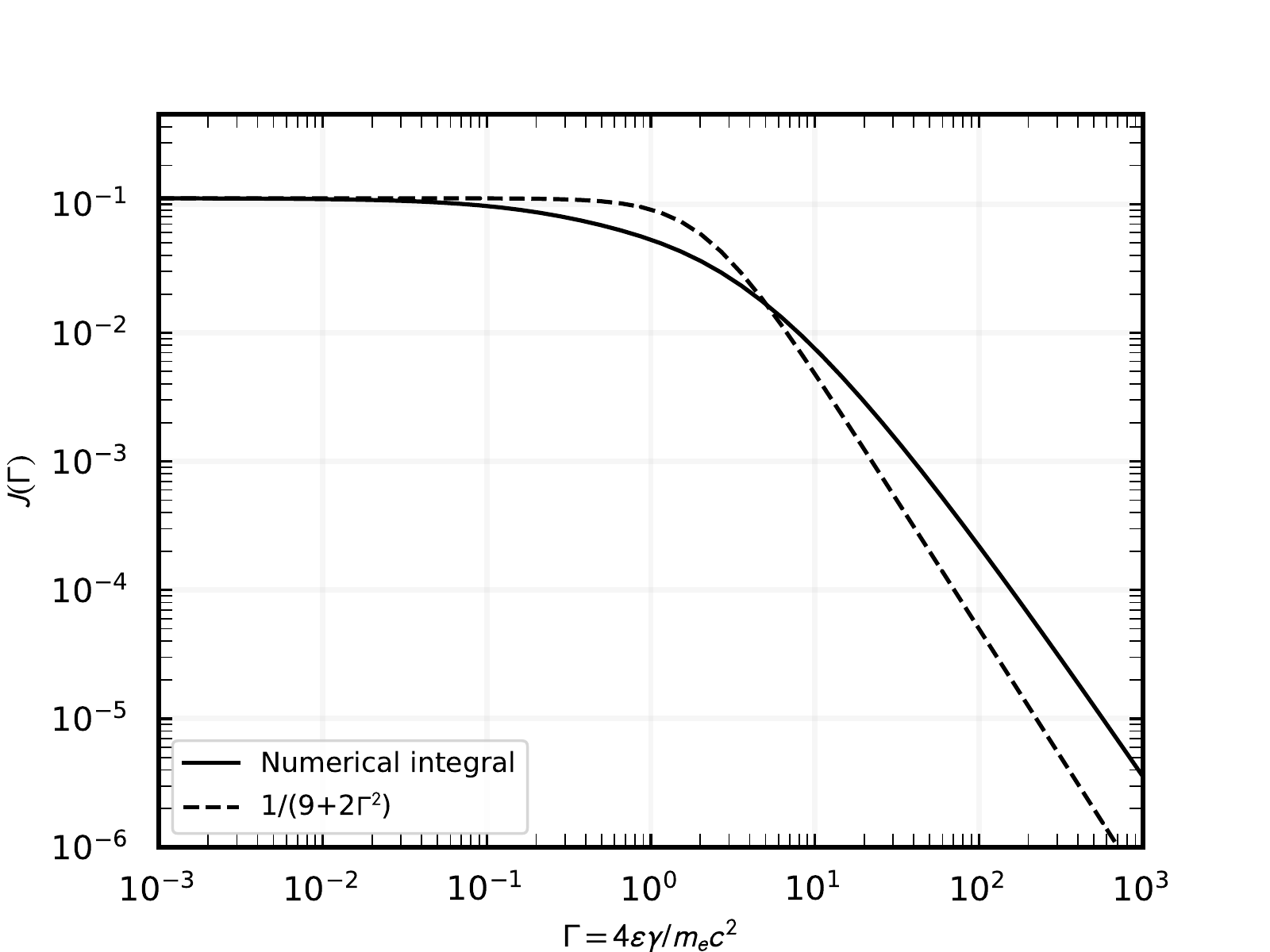}
\includegraphics[width=0.49\textwidth]{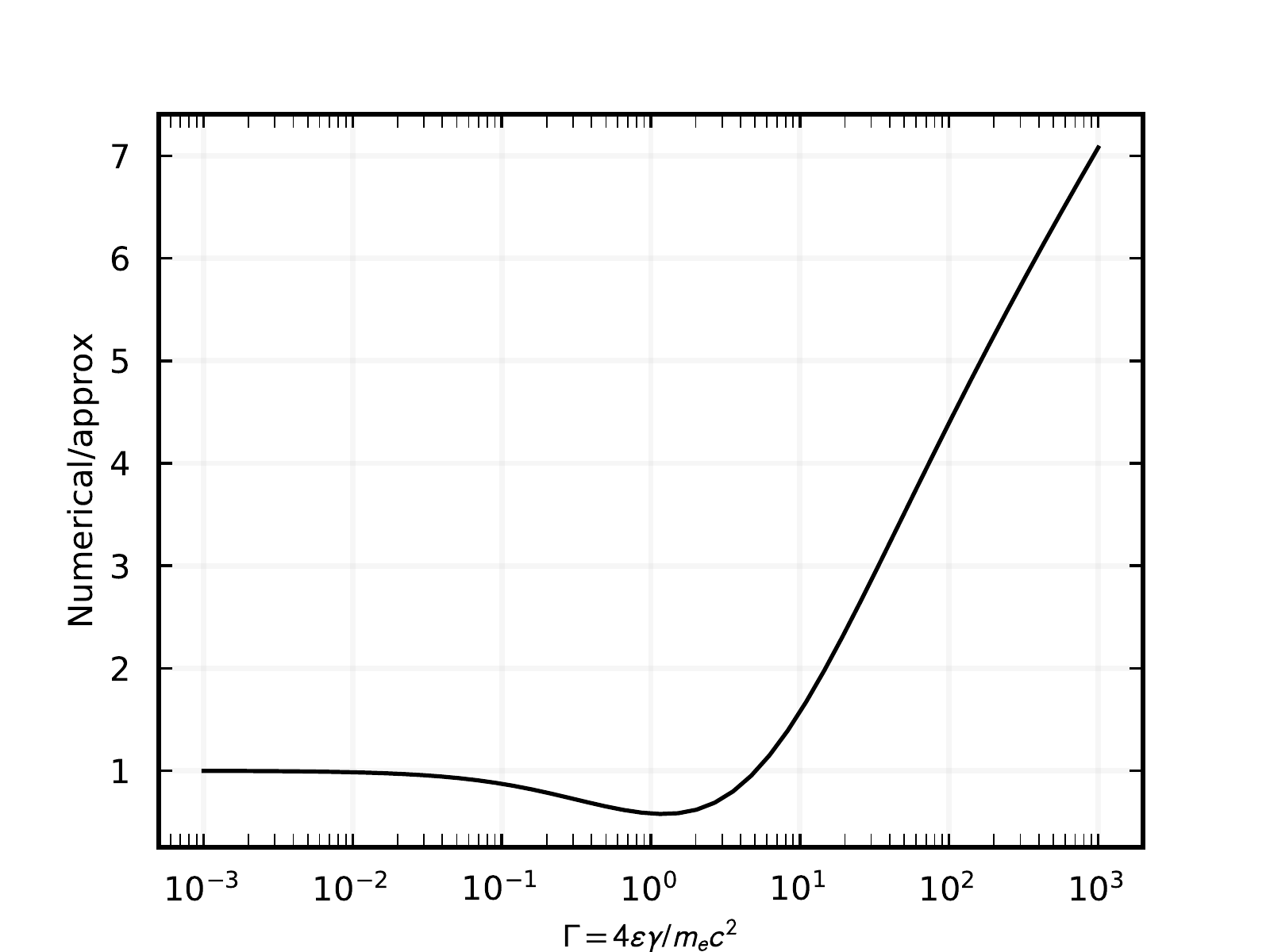}
  \caption{The left (right) panel shows the comparison (ratio) between the numerical calculation of $\mathcal{J}(\Gamma)$ and the approximation made in \cite{2010NJPh...12c3044S} as a function of $\Gamma$.}
  \label{fig:JGamma}
\end{figure*} 

The first and most relevant approximation concerns the integral $\mathcal{J}(\Gamma)$ (see Eq.~\ref{eq:losses} and \ref{eq:G}) that in Ref.~\cite{2010NJPh...12c3044S} is approximated as $J(\Gamma)\approx 1/(9+2\Gamma^2)$.
We show in Fig.~\ref{fig:JGamma} the comparison between this approximated result and the complete numerical calculation.
The approximated function for $\mathcal{J}(\Gamma)$ works well only for $\Gamma<0.1$ and at $\Gamma\approx 3$. In the remaining range of $\Gamma$, the parametrization with $1/(9+2\Gamma^2)$
can reduce the value of $\mathcal{J}(\Gamma)$ with respect to the full numerical computation by a factor of 0.5 for $\Gamma$ around 1, and increase it of a factor between 2 and 7 for $\Gamma = [2,10^3]$.
In particular, we verified that the ratio between $1/(9+2\Gamma^2)$ and the numerical integral is equal to 1 within $20\%$ only for $\Gamma<0.1$ and at $\Gamma\approx 3$. In the remaining range of  $\Gamma$, this substitution can introduce a difference between the numerical and approximated computation of a factor of 2 for  $\Gamma$ around 1, and even 2-3 for $\Gamma>10$.
Considering the starlight ISRF at the peak of its energy distribution, i.e.~$\epsilon\sim 1$ eV, $\Gamma > 2$ implies an electron energy  $E>130$ GeV. These energies for $e^-$ are relevant for interpreting AMS-02 data. 

This discrepancy has important repercussions on the energy loss calculation, since the function $J(\Gamma)$ enters in the the total ICS energy loss rate.
For example, considering the starlight component of the ISRF, peaked at about $\sim 12000$~K, we expect that the {\tt ICS approx} should deviate significantly from the {\tt ICS numerical} for $\Gamma>0.1$, which translates to an electron energy above 5 GeV, in rough agreement to what is observed in Fig.~\ref{fig2:losses} by comparing the blue solid line ({\tt ICS numerical}) to the dotted black line ({\tt ICS approx}).
These conclusions do not depend on the ISRF modeling, as we find very similar results when using also the models reported in \cite{Porter_2006,2010A&A...524A..51D}.

Once the approximated value for the integral $\mathcal{J}(\Gamma)$  is introduced, a second approximation is required in order to provide the final analytical solution. Following the steps in Ref.~\cite{2010NJPh...12c3044S}, the integral over the photon energy is parameterized as a function of the variable $A= \frac{3 m_e c^2}{\sqrt{32} K_B T \gamma}$. If the leading contributions for both regimes of small and large $A$ are considered, an approximation of at most a factor of two for $A\sim1$ is further introduced.

\section{Results with GALPROP }\label{sec:appendix_galprop}
In order to check a possible influence of convective winds, diffusive reacceleration and other low-energy effects on the main conclusions of the paper, we solve the transport equation for the secondaries, the SNR and PWN contributions using the last version of GALPROP v56  \footnote{\url{https://galprop.stanford.edu/}}. We refer to \cite{Strong:2007nh,Strong:1998fr,Strong:2015zva} for a detailed description of the code.  
In this model, the convection, reacceleration and the spatially dependendent energy losses are taken into account by numerically solving the complete transport equation. GALPROP is used in its standard configuration, solving the transport equation in a 2D grid with spacing dz=0.1kpc and dr=1kpc. Energy losses associated to inverse Compton (computed with full Klein-Nishina ICS formalism on ISRF Porter2006), synchrotron emission (on Galactic magnetic field of $B_0=3\mu$G), and bremmshtralung are taken into account.  
Propagation parameters are taken as the BASE +inj+va model,  as recently fitted using GALPROP on recent cosmic-ray nuclei data in \cite{Korsmeier:2021brc}. In particular, in this model a non-zero value for the convection ($v_{\rm 0conv}=5.02$~km s$^{-1}$)  and Alfven speed  ($v_{\rm a}=10.68$~km s$^{-1}$) are found. The modeling of the injection spectrum for both the SNR and PWN strictly follows what done with the semi-analytical model in Sec.\ref{sec:model}.

\end{document}